\documentclass[%
reprint,
aps,
]{revtex4-1}
\usepackage{graphicx}
\usepackage{hyperref}

\usepackage{color}
\usepackage{multirow}
\usepackage{dcolumn}
\usepackage{bm}
\usepackage{amsmath}
\usepackage{natbib}
\usepackage{docmute}
\usepackage{amssymb}
\usepackage{ulem}

\begin{document}


\title{Time-Reversal Invariant Superconductivity of  Sr$_{2}$RuO$_{4}$ Revealed by Josephson Effects}


\author{Satoshi Kashiwaya}
\affiliation{Department of Applied Physics, Nagoya University, Nagoya 464-8603, Japan}
\author{Kohta Saitoh}
\altaffiliation{current affiliation: Nanophoton Corporation, Osaka 565-0871, Japan}
\author{Hiromi Kashiwaya}
\author{Masao Koyanagi}

\affiliation{National Institute of Advanced Industrial Science and Technology (AIST), Tsukuba, Ibaraki 305-8568, Japan}

\author{Masatoshi Sato}
\affiliation{Yukawa Institute for Theoretical Physics, Kyoto University, Kyoto 606-8502, Japan}

\author{Keiji Yada}
\affiliation{Department of Applied Physics, Nagoya University, Nagoya 464-8603, Japan}

\author{Yukio Tanaka}
\affiliation{Department of Applied Physics, Nagoya University, Nagoya 464-8603, Japan}

\author{Yoshiteru Maeno}
\affiliation{Department of Physics, Kyoto University, Kyoto 606-8502, Japan}

\date{\today}

\begin{abstract}
Sr$_{2}$RuO$_{4}$ is one of the most promising candidates of a topological superconductor with broken time-reversal symmetry, because a number of experiments have revealed evidences for a spin-triplet chiral $p$-wave superconductivity. In order to clarify the time-reversal symmetry of Sr$_{2}$RuO$_{4}$, we introduce a novel test that examines the invariance of the Josephson critical current under the inversion of both the current and magnetic fields, in contrast to the detection of a spontaneous magnetic field employed in past experiments. Analyses of the transport properties of the planar and corner Josephson junctions formed between Sr$_{2}$RuO$_{4}$ and Nb reveal the time-reversal invariant superconductivity, most probably helical $p$-wave, of Sr$_{2}$RuO$_{4}$. This state corresponds to a yet-to-be confirmed $topological crystalline superconductivity$ that can host two Majorana edge modes at the surface protected by crystalline mirror symmetry.
\end{abstract}


\maketitle

\section{Introduction}
Spontaneous symmetry breaking is one of the fundamental concepts in nature. The electrons in condensed matter transit to a lower energy state with lower symmetry when the temperature is reduced. A typical example is the Bardeen-Cooper-Schrieffer (BCS) type superconductor (SC), in which the electrons transit to the superconducting state accompanied by gauge symmetry breaking ~\cite{1}. Recently, unconventional non-BCS superconducting states with additional spontaneous symmetry breaking have been explored in various novel SCs. The unusual superconducting state of layered perovskite Sr$_{2}$RuO$_{4}$ (Fig. 1A) with a critical temperature ($T_c$) of 1.5 K~\cite{2} has been a topic of intense debate over the last two decades because of numerous experimental results suggesting the first spin-triplet, time-reversal symmetry (TRS) breaking superconductivity albeit with unresolved issues~\cite{3,4,5,6,7}.

Measurements of the Knight shift exhibiting totally different behavior from that of spin singlet SCs strongly support spin triplet superconductivity~\cite{4,8}. The interference patterns of SQUID suggest the odd parity pairing of Sr$_{2}$RuO$_{4}$~\cite{9}. Accepting the spin triplet superconductivity, a crystal structure with $D_{4h}$ symmetry allows for six possible triplet pairing states under a quasi-two-dimensional (2D) Fermi surface. They are classified into two classes: two chiral states that break TRS with a $d$-vector aligned to the $c$-axis (Fig. 1D), and four helical states that preserve TRS but break the spin-orbit symmetry with a $d$-vector lying in the plane (Fig. 1B, 1C) ~\cite{3}. The six states are typical examples of 2D topological SCs characterized by gapless edge-state formation~\cite{4,10}. In fact, the formation of the edge states has been observed as broad zero-bias conductance peaks of tunnel junctions~\cite{11,12, 50}. Because of such unique characteristics, the final identification of the pairing symmetry is an outstanding current issue in superconductivity research.

The key issue in establishing the pairing states of Sr$_{2}$RuO$_{4}$ is the presence or absence of TRS. Since a recent theory clarifies the competing energy levels of the chiral and helical states~\cite{13}, an experimental determination is strongly desired. Among several past experiments that tested TRS, an increase in the muon spin relaxation rate owing to a spontaneous magnetic field~\cite{14} and the presence of a finite Kerr rotation in the magneto-optic Kerr effect~\cite{15} suggest a broken TRS. Based on these results, chiral $p$-wave superconductivity has been widely accepted until recently. However, real-space detections of the spontaneous magnetic field originating from the chiral edge current~\cite{16,16-2} were unsuccessful with scanning SQUIDs~\cite{17,18} and scanning Hall probes~\cite{19}. Although the compatibility of the chiral superconductivity with the lack of a spontaneous edge current has been theoretically discussed~\cite{5,20,40,41,42,43,44,45}, the origin of the inconsistency has not been fully resolved yet. Therefore, the unambiguous determination of TRS based on a new reliable experimental probe is strongly desired.

In this study, we present strong evidence for a time-reversal invariance of superconductivity of Sr$_{2}$RuO$_{4}$ using Sr$_{2}$RuO$_{4}$/Nb Josephson junctions (JJs). The significant influence of dynamical domain motion due to multi-component superconductivity in the JJ characteristics has been known~\cite{21}. Such influence was successfully excluded by reducing the junction size~\cite{22}. Here, we introduce a novel current-field inversion (CFI) test of TRS that examines the invariance of the Josephson critical current under the inversion of both the current and magnetic fields. By introducing the CFI test, one can clearly resolve the harmonic components of the Josephson current, thereby identify TRS~\cite{23,24,64}. We accumulated critical current data of both planar and corner JJs and for all combinations of the positive and negative directions of the current and field. Based on a symmetry analysis of the critical current-magnetic field ($I_C-H$) patterns together with other basic transport properties of Sr$_{2}$RuO$_{4}$/Nb JJs, we obtain the convincing conclusion that Sr$_{2}$RuO$_{4}$ is a TRS invariant helical triplet SC. This suggests the realization of a novel “topological crystalline superconductivity” with stable Majorana edge modes at the surface of Sr$_{2}$RuO$_{4}$ (Fig. 1E)~\cite{25}.
\par
\begin{figure}[t]
  \begin{center}
		\includegraphics[width = 8.5cm]{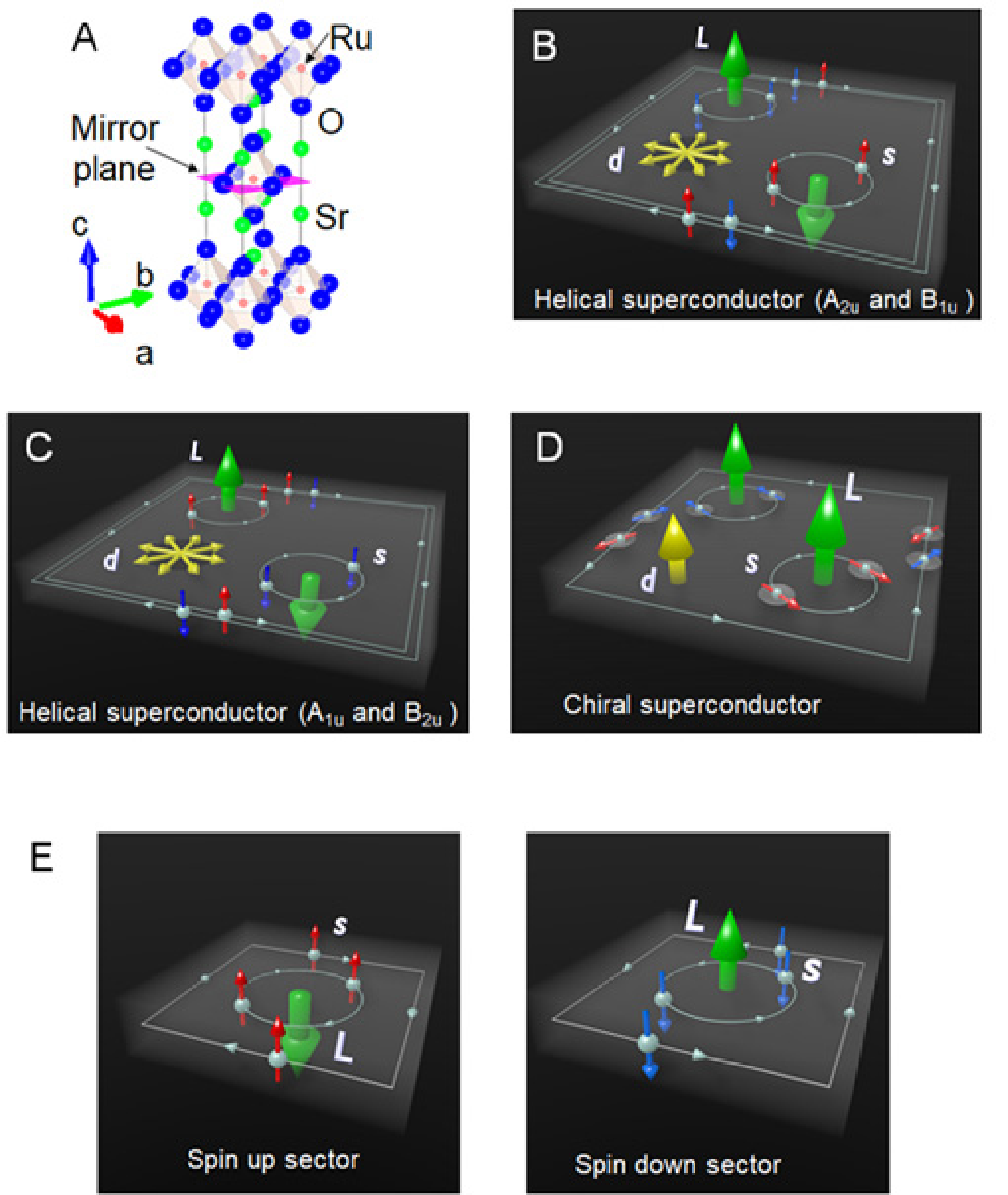}
  \end{center}
  \caption{Spin-triplet Cooper pairs and the topological crystalline superconductor. (A) Crystal structure and the mirror plane of Sr$_{2}$RuO$_{4}$. Green, red and blue spheres represent Sr, Ru and O atoms, respectively. (B) Illustration of spin-triplet Cooper pairs and corresponding edge currents on the basal $ab$-plane for a set of helical states in which orbital angular momentum and spins are anti-parallel ($A_{2u}$ and $B_{1u}$ in Mulliken notation). Red and blue arrows represent up and down Cooper-pair spins (s). Yellow arrows represent the $d$-vector of spin-triplet superconductivity, and green arrows the orbital angular momentum (L). Shown in light blue lines are helical edge states carrying pure spin current without charge current. Two distinct helical states $A_{2u}$ and $B_{1u}$ differ by phase difference between spin sectors, 0 or $\pi$. (C) Illustration of another set of helical states ($A_{1u}$ and $B_{2u}$) superconductors in which orbital angular momentum and spins are antiparallel. Two distinct helical states $A_{1u}$ and $B_{2u}$ differ by phase difference between two spin sectors, 0 or $\pi$. (D) Illustration of chiral states ($E_{u}$). Edge currents for the two spin states are in the same direction for the chiral state, whereas they are in opposite directions for the helical state. (E) Spin-up and spin-down sectors of the helical state. The helical state under mirror symmetry leads to topological crystalline superconducting state with a stable Majorana zero mode for each spin sector.}
  \label{Fig1}
\end{figure}
%
%
\section{Theoretical background}
First, we describe the basic concept of the test for the TRS using Josephson effect employed in the present study. We assume a JJ between composed of a conventional SC (CSC) and an unconventional SC (USC). The current phase relation of the Josephson current $I(\varphi)$ ($\varphi$:phase difference between the two SCs) can generally be decomposed into harmonic terms:

\begin{align}
    I(\varphi) = \sum_{n} \left\{ I_n^s\sin(n\varphi)+I_n^c\cos(n\varphi) \right\},
\end{align}
where $n$ is a positive integer.
When the USC preserves TRS, only the sine terms becomes non-zero, whereas when the USC breaks TRS cosine terms also become finite~\cite{26}. Therefore, the determination of TRS based on the Josephson effect is equivalent to identifying the presence of cosine terms in Josephson current components. Moreover, we must consider the effects of spin-triplet pairing of the USC: the first terms ($n$ = 1) disappear without the spin-orbit (SO) interaction owing to the spin space orthogonality between the singlet and triplet states. In contrast, in the presence of SO interaction, these first-order terms recover non-zero values due to spin-flip scattering at the interface and/or in the bulk. Nevertheless the amplitude of the first-order terms tend to be suppressed due to the spin space orthogonality compared to those of a JJ composed only of CSCs.
\par
\begin{figure*}[t]
  \begin{center}
		\includegraphics[width=1.0\linewidth]{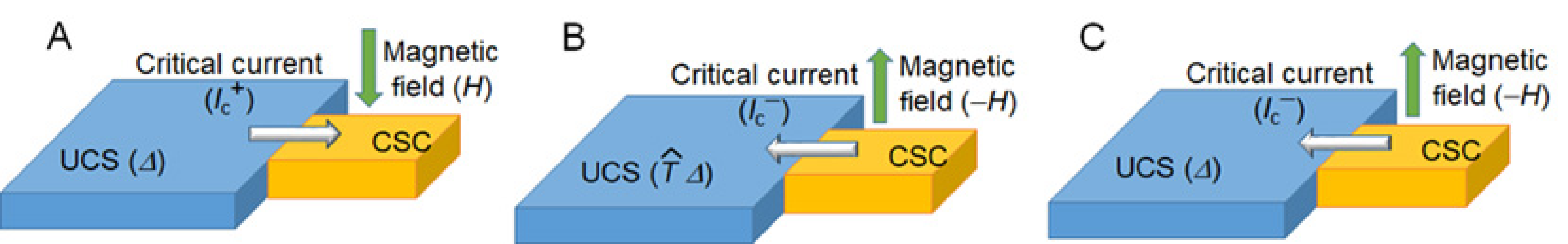}
  \end{center}
  \caption{(A) Illustration of current-biased Josephson junction of conventional superconductor (CSC)/unconventional superconductor (USC) with order parameter $\Delta$. Critical current is represented by $I_C^+$ in an applied magnetic field $H$. (B) Time-reversal system obtained by reversing current and magnetic field directions, as well as time reversal of order parameter (($\hat{T}\Delta$); $\hat{T}$ is time reversal operator). Amplitude of critical current represented by $I_C^-(-H)$ is equivalent to $I_C^+(H)$ in（A）. (C) Experimentally feasible system of reversed current and reversed magnetic field directions. Since order parameter cannot be tuned externally, it is unchanged from (A), $-I_C^-(-H)$ is inequivalent to $I_C^+(H)$ unless USC is invariant to time-reversal. Therefore, trivial equivalence between $I_C^+$ and $-I_C^-(-H)$ disappears for USC with broken TRS owing to $\hat{T}\Delta\neq \Delta$.
}
  \label{Fig2}
\end{figure*}
Kawai $et$ $al$. calculated the Josephson current by taking into account of realistic multiple-band structure of Sr$_{2}$RuO$_{4}$ and the SO interaction at the interface~\cite{49,27}. According to their results, which are summarized in Table I and II of ~\cite{27}, only the $I_n^s$s (sine terms) become non-zero for the helical states, and the corresponding $I(\varphi)$ is an odd function $I(\varphi)=-I(-\varphi)$] similar to that of JJs composed only of CSCs. By contrast, for the chiral state, some values of $I_n^c$ (cosine terms) become nonzero, and the corresponding $I(\varphi)$ is no longer an odd function as a consequence of the broken TRS. Therefore, the helical states and the chiral states can be empirically discriminated by examining the presence of cosine terms appeared in $I_C$-$H$ patterns in both the planer and the corner Josephson junction. 
\par
To identify the presence of cosine terms, the current and magnetic field inversion test for the TRS is performed on the magnetic field ($H$) response of $I_C$ ($I_C$-$H$ pattern) of the JJs. 
Different from conventional analysis, we explicitly examine the critical current $I_C$ for both the positive ($I_C^+> 0$) and negative ($I_C^- < 0$) directions. We consider three types of symmetry of the$I_C$-$H$ pattern (see Fig. 5A): (i) current inversion [CI, $I_C^+(H) = -I_C^-(H)$], (ii) magnetic field inversion [FI, $I_C^\pm(H) =I_C^\pm(-H)$], and (iii) current and magnetic field inversion [CFI, $I_C^\pm(H) = -I_C^\mp(-H)$].
Since the presence of the first and second harmonic terms was identified by Shapiro steps, three candidates can be listed as the pairing states: (I) helical SC [$I(\varphi)= I_1^s \sin\varphi + I_2^s \sin2\varphi$], (II) single-band chiral SC with SO interaction [$I(\varphi)= I_1^c\cos\varphi + I_2^s \sin2\varphi$], and (III) multiband chiral with SO interactions [$I(\varphi)= I_1^s\sin\varphi + I_1^c\cos\varphi + I_2^s\sin2\varphi+ I_2^c\cos2\varphi$]. The relation between the symmetry of $I_C$-$H$ and the above candidates is summarized in Table I. The influence of extrinsic effects such as the self-field and the non-uniform current distribution are also taken into account. It is important to note that testing the CFI symmetry of both the planar and corner JJs is essential in discriminating between these three candidate superconducting states. 
\par
The essence of physical concept is desicribed in order to intuitively understand the underlying physics. We assume a current-biased JJ between CSC and UCS with the order parameter $\Delta$, as shown in Fig. 2A. Time-reversal symmetry in the junction components except in the USC, as well as inversion symmetry of the USC, are assumed. Figure 2B shows the time reversal of Fig. 2A and has an equivalent Ic with the opposite sign. The configuration shown in Fig. 2B is obtained from Fig. 2A with three operations: i) reversing the current direction, ii) reversing the magnetic field, and iii) time reversing the UCS order parameter  $\Delta$ ($\hat{T}\Delta$; $\hat{T}$ ̂ is the time-reversal operator). In the experimentally feasible situation, we can externally control only the current (current inversion, CI) and the magnetic field directions (field inversion FI), while the superconducting order parameter stays unchanged from $\Delta$, as shown in Fig. 2C.
\par
If the UCS is a time-reversal invariant SC, the relationship of $\hat{T}\Delta=\Delta$ leads to the equivalence between Fig. 2B and Fig. 2C, and thus the CFI invariance $I_C^+(H) = -I_C^-(-H)$ should be preserved. On the other hand, if UCS is a TRS-broken SC ($\hat{T}\Delta\neq \Delta$), the CFI invariance does not hold because Fig. 2B and Fig. 2C are no longer equivalent. Therefore, the TRS and the CFI symmetry of $I_c-H$ are strongly correlated. One exceptional case is the planar junction of a single-band (SB) chiral SC with SO interaction [$I(\varphi) = I_1^c \cos\varphi+ I_2^s \sin2\varphi$]. In this case, $I(\varphi)$ is not an odd function but an antisymmetric function with respect to $\varphi= \pm\pi/2$ [$I(\varphi \pm \pi/2) = -I(-\varphi \mp \pi/2)$]. As a result, $I_C-H$ retains the CFI symmetry even though USC is time-reversal broken, as summarized in Table I.
\par
	It is noted that the CFI symmetry is insensitive to the presence of extrinsic effects such as SO interactions, non-uniformity of the current distribution, and the self-field effect. Therefore the present test is quite robust against the experimental difficulties. In fact, $I_C$ peaks tend to shift to a finite magnetic field due to the influence of the self-field in the past experiments ~\cite{24}. Temperature dependence of the peak position need to be precisely measured to exclude the influence of the self-magnetic field. Such a method cannot be applied to superconductors whose pairing symmetries can be varied depending on the temperature. Whereas, the present test has an advantage that the TRS can be judged based on a single temperature data because it is intrinsically insensitive to the self-field. Furthermore, the test based on the CFI invariance does not rely on the detection of a magnetic field induced by the edge current~\cite{16}, which is not topologically protected and may be substantially weakened by various effects ~\cite{40,41,42,43,44,45}. Whereas, the present method does not have the ability to discriminate between the two types of helical states shown in Fig. 1B, 1C. In addition, the present test does not properly work if the domain boundaries move during the measurement.
\par
\begin{table}[t]
\caption{Relation between symmetry of $I_C-H$ and pairing states of conventional superconductor (SC)/spin-triplet SC Josephson junction. Different from conventional analysis, we explicitly consider critical current Ics for both positive ($I_C^+ > 0$) and negative ($I_C^- < 0$) directions. We consider three types of symmetry of $I_C-H$: (i) current inversion [CI, $I_C^+(H) = -I_C^-(H)$], (ii) magnetic field inversion [FI, $I_C\pm(H) =I_C\pm(-H)$], and (iii) current and magnetic field inversion [CFI, $I_C^+(H) = -I_C^-(-H)$] (see Fig. S2D). Since presence of first and second terms of Josephson currents have been identified in microwave response, three candidates of pairing states are listed: i) helical with spin-orbit  interaction [$I(\varphi) = I_1^s\sin\varphi+ I_2^s\sin2\varphi]$, ii) single-band (SB) chiral with SO [$I(\varphi) = I_1^c\cos\varphi+ I_2^s\sin2\varphi$], and iii) multiband (MB) chiral with spin-orbit  interaction [$I(\varphi) = I_1^s\sin\varphi+ I_1^c\cos\varphi+ I_2^s\sin2\varphi+ I_2^c\cos2\varphi$]. For reference, case of a conventional spin singlet SC instead of spin triplet SC is also included. Among these pairing states, conventional and helical preserve the time-reversal symmetry, while SB chiral and MB chiral break time-reversal symmetry. Mark S(A) denotes symmetry (asymmetry) with respect to inversion. To certify consistency with actual experiments, influences of current non-uniformity and self-magnetic field are also considered. Table claims that three types can be discriminated by testing CFI symmetry in planar and corner Josephson junctions.
} 
 \begin{tabular}{|c|c|c||c|c|c|} \hline
    Pairing states &Type &Inhomo\&SF&CI&FI& CFI  \\ \hline 
   &Planer &$-$& S& S &  S \\ \cline{3-6}
      Conventional&       &$\surd$& A& A &  S \\ \cline{2-6}
        Helical&  Corner     &$-$&S& S &  S \\ \cline{3-6}
       &       &$\surd$& A& A &  S \\ \cline{1-6}
   &Planer &$-$& S& S &  S \\ \cline{3-6}
        SB chiral &       &$\surd$& A& A &  S \\ \cline{2-6}
      &  Corner     &$-$& S& A &   A \\ \cline{3-6}
      &       &$\surd$& A& A &  A \\ \hline
         &Planer &$-$& A& S &  A \\ \cline{3-6}
        MB chiral &       &$\surd$& A& A &  A \\ \cline{2-6}
     &  Corner     &$-$& A& A &   A \\ \cline{3-6}
      &       &$\surd$& A& A &  A \\ \hline
    \end{tabular}
  \label{table}
\end{table}
\section{Experimental}
The Sr$_{2}$RuO$_{4}$/Nb Josephson junction (JJ) interface needs to be formed at the surface perpendicular to the $ab$-plane of Sr$_{2}$RuO$_{4}$ in order to detect the internal phase of the superconductivity, as shown in Fig. 3A~\cite{28}. Single crystals of Sr$_{2}$RuO$_{4}$ grown by a floating zone method~\cite{31} are polished to plates of several microns in thickness prior to the deposition of Nb. Since the superconductivity at cleaved surfaces is easily degraded against atmospheric exposure, we developed a process to crash the plates in a vacuum and subsequently deposit the counter electrodes in-situ. To fabricate junctions in a 3D structure, patterning of four-terminal electrodes (Fig. 3B) is carried out by a focused ion beam (FIB) process. 

The FIB process has an advantage in that the junction size can be changed successively even after measuring the transport at low temperatures. More details on the fabrication process of the JJs are presented in reference~\cite{28}. The transport measurements are conducted in a conventional four-terminal configuration using a handmade small superconducting magnet. The bias voltage is generated with a waveform generator (Agilent 33521A), and the output signal is amplified with an input coil (NF LI-772N) and a lock-in amplifier (NF LI575). The residual magnetic field of the measurement system is confirmed to be less than 4 mOe owing to the magnetic field shielding located both at room temperature and at low temperature ~\cite{32}.The empirical microwave responses of the JJs are measured by using a loop antenna consisting of a CuNi wire connected to a function generator (HP 8672D) via a coaxial line. 
\begin{figure}[t]
  \begin{center}
		\includegraphics[width=1.0\linewidth]{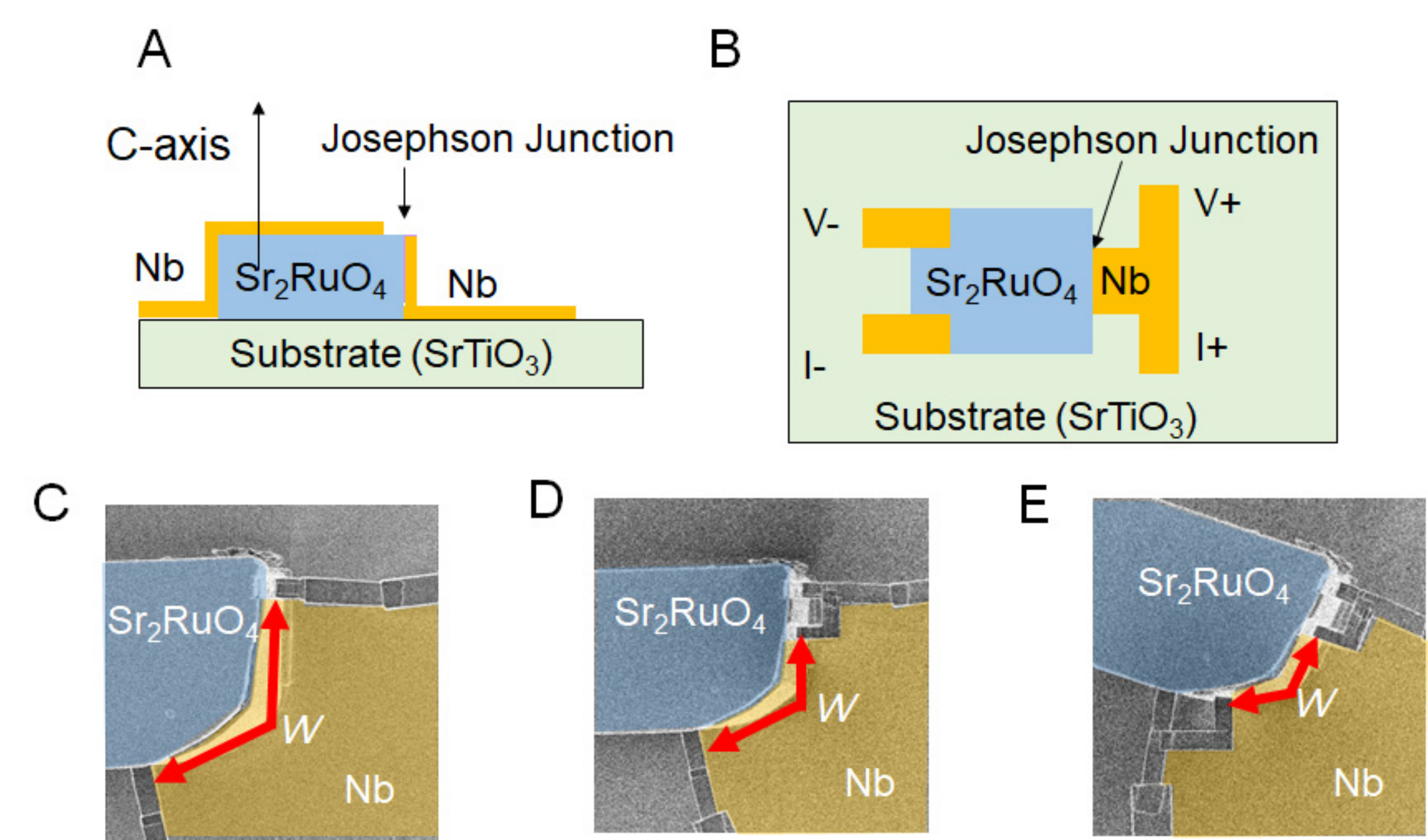}
  \end{center}
  \caption{Geometry and scanning ion microscopy images of Nb/Sr$_{2}$RuO$_{4}$ corner junction testing the current uniformity.
(A) Schematic cross section of present Josephson junctions. (B) Schematic illustration of junction configuration. (C, D, E) Scanning ion microscopy (SIM) images of corner Junction Y at each stage of successive reduction of junction width. Thicknesses of Nb and Sr$_{2}$RuO$_{4}$ are 100 nm and 10 $\mu$m, respectively. Current uniformity was confirmed by systematic change of $I_C$ in accordance with junction size $w$ [C, $w$ = 18 $\mu$m (Ic = 8 mA); D, $w$ = 14 $\mu$m (Ic = 5.7 mA), and E, $w$ = 10 $\mu$m (Ic = 4.5 mA)] 
}
  \label{Fig3}
\end{figure}
%
%
%
\par
	As a background in the present work, we should point out that the intrinsic transport properties of the JJs composed of Sr$_{2}$RuO$_{4}$ are still not wholly clarified in previous works. This is owing to the serious influences of dynamical domain motions in $I_C-H$ ~\cite{21,22}. If these superconducting domain boundaries existing in the junction begin to move during the $I_C-H$ measurements, the correspondence of $I_C$ for different $H$ is lost because Ic depends on the domain texture ~\cite{34,35}. This effect results in the hysteretic $I_C-H$ patterns reported by~\cite{21}. Similar anomalies owing to the domain dynamics were reported in other studies ~\cite{22,30,36,37}. In the present experiments, most of the fabricated JJs show a large variation in the $I_C-H$ patterns when the junction size is larger than dozens of micrometers. 
	We found that $I_C-H$ changes to hysteretic patterns similar to those reported by~\cite{21} by miniaturizing the junction size to a few tens of microns. By further decreasing the size, the $I_C-H$ patterns converge to stable patterns~\cite{22}. It is noted that such systematic variation is consistent with the multicomponent SC: the presence of dynamical superconducting domains with sizes of several microns modifies the transport properties. We believe that the effect of dynamical domain motion is successfully excluded in the present results because the observation of current-field inversion (CFI) symmetry ensures that the domains are stable within the corresponding magnetic field range even if they exist. This is because the consistency between $I_C^+(H)$ and $-I_C^-(-H)$ would have been lost if the domains move during the measurement.
\par
\begin{figure}[t]
  \begin{center}
		\includegraphics[width=1.0\linewidth]{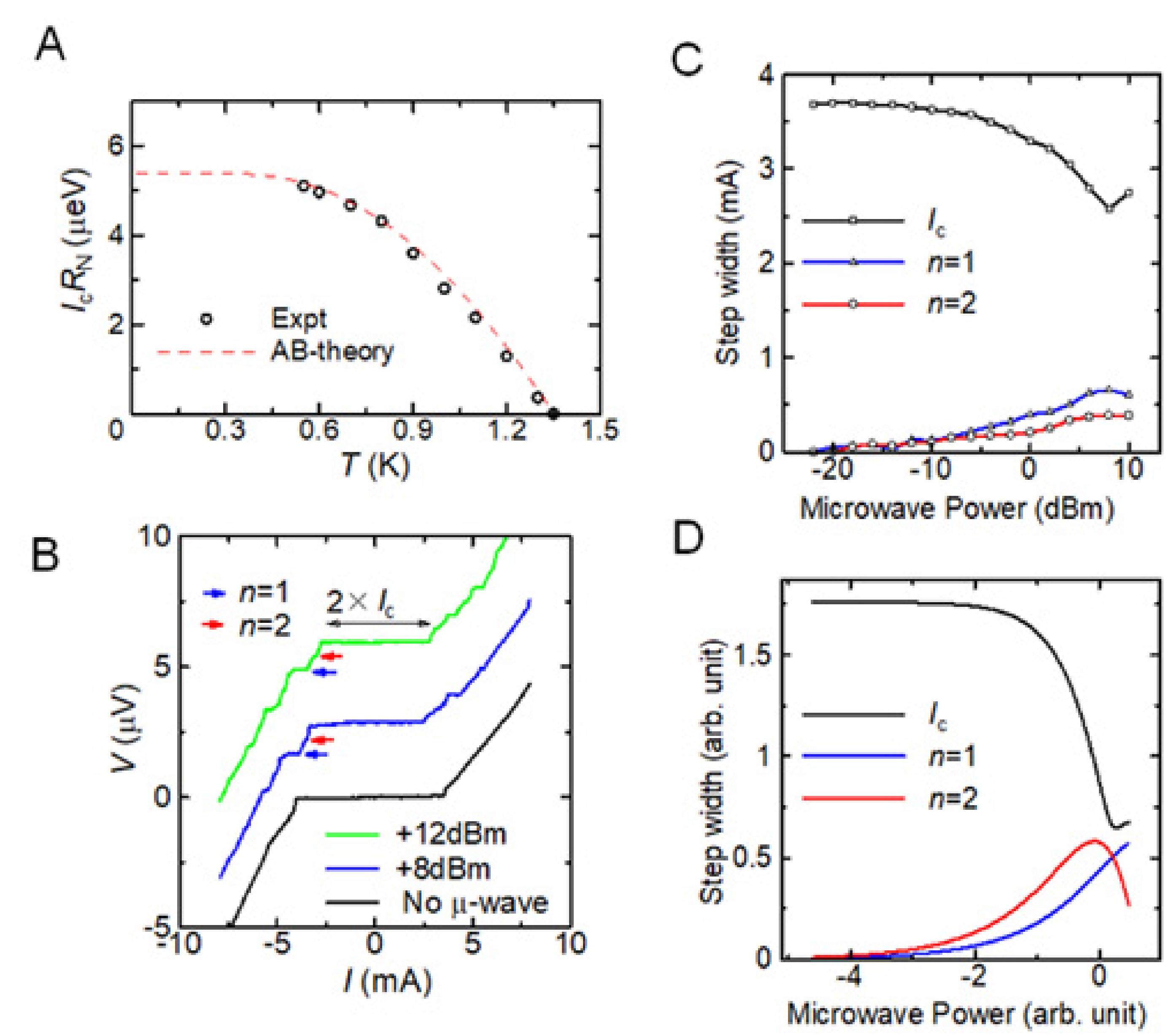}
  \end{center}
  \caption{Transport properties of Sr$_{2}$RuO$_{4}$ in Sr$_{2}$RuO$_{4}$/Nb Josephson junctions.
(A) Temperature dependence of the product of the critical current and normal state junction resistance ($I_CR_N$) of Junction Y (open circles). The curves are well fitted with Ambegaokar-Barratof theory for a typical Josephson tunneling junction, although its amplitude is two orders of magnitude smaller than expected. (B) Current-voltage ($I-V$) curves of Junction Y measured at 0.8 K with and without microwave irradiation at frequency f = 500 MHz. Black arrow indicates 2Ic. Blue and red arrows represent Shapiro steps corresponding to the first ($n = 1$) and second ($n = 2$) harmonic terms. (C) Microwave power dependence of Ic, and step widths for the first and second branches. Amplitude of $I_C$ exhibits anomalous dip at high microwave power. (D) Theoretically calculated microwave power dependence of step widths. Black, blue, and red curves correspond to $I_C$, the first, and second branches, respectively. A dip of power dependence of Ic appears when the first and second harmonic terms coexist in comparable amplitudes.
}
  \label{Fig4}
\end{figure}
%
%
\section{Analysis of $I_c-H$}
We examine two types of JJs: Junction X is a planar junction of which an Nb electrode is formed on a single edge of the crystal; Junction Y is a corner junction formed across two edges. Figures 4A-C show the results of the transport properties of Junction Y. The temperature dependence of the $I_cR_N$ product ($R_N$: the junction resistance just above $T_C$) shown in Fig. 4A mostly follows the Ambegaokar-Baratoff formula~\cite{29} as indicated by the dotted line, which suggests that the present junction is in the tunneling regime.
\par
The amplitude of the $I_cR_N$ product, however, is about two orders of magnitude smaller than the expected value. A similar reduction in $I_cR_N$ is commonly observed in other Sr$_{2}$RuO$_{4}$/Nb junctions. 
Figure 4B shows the microwave response of the current-voltage ($I$-$V$) characteristic of Junction Y. The harmonic terms of the Josephson current can be resolved based on the constant voltage steps (Shapiro steps) under microwave irradiation. In addition to the steps corresponding to the first term ($n$ = 1, $V$ = $hf/2e$), the steps corresponding to the second term ($n$ = 2, $V = hf/4e$) are clearly observed. In the microwave power dependence of the step widths shown in Fig. 4C, the step widths corresponding to the first and second terms exhibit conventional Bessel-function responses, whereas the zero-th branch width (equivalent to $I_C$) shows unconventional power dependence with an unusual minimum.
\par
%
To clarify the origin of uncoventional power dependence of the step widths shown in Fig. 4D,  theoretical curves are calculated based on a conventional RF-driven voltage-bias model~\cite{33}. The microwave power dependence of the steps shown in Fig. 4D is calculated by assuming $I(\varphi) = I_1^s\sin \varphi + I_2^s\sin 2 \varphi$. The zero-th step (=$I_C$) is given by the absolute value of $I_1^s J_0(p) \sin\varphi_0+ I_2^s J_0(2p) \sin 2\varphi_0$, where $J_k(p)$ is the $k$-th Bessel function, $p$ is the microwave power, and $\varphi_0$ (0 < $\varphi_0$< 2$\pi$) is the phase difference giving the maximum value of Ic. The step widths of the first and second terms are given by the absolute value of $I_1^s J_1(p)$ and the absolute value of $I_2^s J_1(2p)$, respectively. We assume $I_1^s \sim I_2^s \sim 1$ in the simulation for simplicity. Indeed, the experimentally detected non-monotonous temperature dependence of Ic suggests that the first and second terms coexist with almost the same amplitudes.
Assuming a JJ of CSC/CSC in the tunneling regime, this is quite anomalous because it requires that the first-order terms are severely suppressed. On the other hand, this is exactly what is anticipated for triplet SC/CSC JJs because the orthogonality of the spin space suppresses the first term and thus $I_cR_N$.
\par 
\begin{figure}[t]
  \begin{center}
		\includegraphics[width=1.0\linewidth]{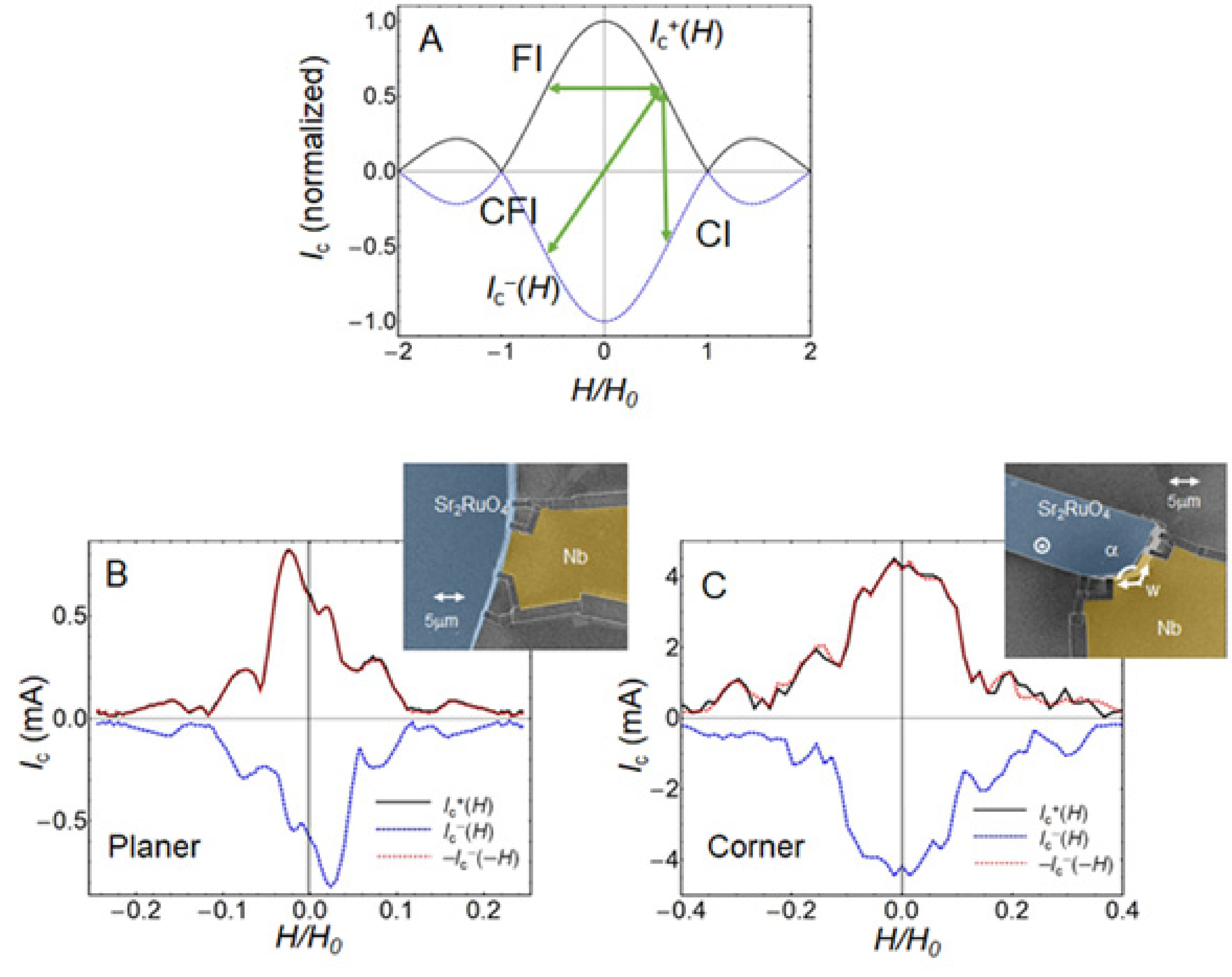}
  \end{center}
  \caption{Time-reversal invariance revealed by current and magnetic field inversion (CFI) of $I_C-H$.
(A) Three types of symmetries in magnetic field dependence of critical current ($I_C-H$) patterns being considered: current inversion (CI), magnetic field inversion (FI), and current and magnetic field inversion (CFI). 
Time-reversal symmetry of a superconductor can be tested by the invariance to the CFI symmetry. (B, C) $I_C-H$ patterns and scanning ion microscopy (SIM) image of planar Junction X (at 0.31 K, $w$ = 7.1 $\mu$m), and those for corner Junction Y (at 0.32 K, w = 10 (5 + 5) $\mu$m) with the apex angle $\alpha$ set at 2/3$\pi$. Patterns include smoothly connected curves for positive side of $I_C$ ($I_C(H)$: black solid curves) and negative side of $I_C$ ($I_C^+(H)$: blue dotted curves), as well as current and field inverted $I_C$ [$-Ic^-(-H)$: red dotted curves] corresponding to blue curves. Horizontal axes are normalized by $H_0 =12.5$ Oe for X and 8.8Oe for Y, which are given by $\Phi_0 w(\lambda_{SRO} + \lambda_{Nb})/\mu_0 $, where $\Phi_0$ is the flux quantum ($20.7\times10^{-4}$ T ($\mu$m)$^2$), and $\lambda_{SRO}$(=190 nm in the $ab$-plane) and $\lambda_{Nb}$ (=44 nm) are penetration depths in Sr$_{2}$RuO$_{4}$ and Nb, respectively. The period of the oscillation is largely suppressed because of the focusing effect of magnetic field~\cite{28}. The CFI symmetry demonstrated by consistency between black and red curves in both junctions identifies time reversal invariance of superconductivity of Sr$_{2}$RuO$_{4}$.
}
  \label{Fig5}
\end{figure}
%
%
In many previous experiments using Sr$_{2}$RuO$_{4}$, identification of the TRS has been accomplished based on the detection of a spontaneous magnetic field. The spontaneous magnetic field generation at the edge of broken TRS was theoretically predicted by Matsumoto and Sigrist ~\cite{16, 16-2}. Experimentally suggested broken TRS was presented based on $\mu$SR by detecting the finite amplitude of the magnetic field in the bulk ~\cite{14,37-2}, and on the Kerr effect by detecting the magnetization at the surface ~\cite{15}, although the origin of the giant Kerr rotation angle is still controversial ~\cite{38}. On the other hand, trials to detect the magnetic field in real space probes have failed when using a scanning SQUID ~\cite{17,18}, a scanning Hall element ~\cite{19}, and a micro-SQUID ~\cite{39}. Although several theories have explored the compatibility of the chiral superconductivity with a lack of spontaneous edge current ~\cite{20,40,41,42,43,44,45}, the inconsistency has not been resolved yet. Therefore, the determination of a time-reversal symmetry not relying on the detection of the magnetic field is strongly needed.
	While in the present work, we focus on the phase sensitivity of JJs and the symmetry of $I_C-H$ pattern which is quite robust against the experimental details as stated above. In fact, theoretical predictions to detect the phase shift of an odd-parity SC ~\cite{23,64} were applied to the detection of the $d$-wave superconductivity in cuprates~\cite{24,46}. Although the possibility of odd-parity pairing in Sr$_{2}$RuO$_{4}$ was presented using mm-scale JJs by Nelson et al.~\cite{9}, the effects of the phase shift on micrometer-scale domain boundaries~\cite{26} and their dynamics~\cite{34} were not taken into account. Here, we develop this idea to test the TRS of novel SCs through the symmetry of $I_C-H$ of JJs between CSCs and UCSs.
\par
	The uniformity of $I_C$ in the junction is another important factor in validating the corner junction results to exclude the possibility that the Josephson current flows only through a single wing of the corner juction. The uniformity at a scale of a few microns has been confirmed by sequentially miniaturizing the junction size by using FIB. Figures 3C–E show scanning ion microscope (SIM) images of Junction Y at each stage of the successive reductions of the junction size. With a decrease in the junction width $w$, $I_C$ systematically changed from 8.0 mA (w = 18 $\mu$m), 5.7 mA (w = 14 $\mu$m), then to 4.5 mA (w = 10 $\mu$m). Since the amplitude of $I_C$ is almost proportional to $w$, the possibility that the current is concentrated on a single edge of the corner junction can be rejected. Nevertheless, a current uniformity smaller than the micrometer scale has not been evaluated. This is not a serious problem for the interpretation because the presence of small-scale non-uniformity does not change the conclusion of the present paper.
\par
The experimental results of $I_C$-$H$ are examined by referring to Table SI. Figures 5B and 5C show SIM images and $I_C$-$H$ patterns of the planar and corner JJs. The sizes of the junctions are selected so that the dynamical domain motion can be excluded by referring to the results in Ref~\cite{22}. In addition to $I_C^+(H)$ (black) and $I_C^-(H)$ (blue), the inverted $-I_C^-(-H)$ (red) is plotted to certify the CFI symmetry. It is quite clear that the intricate $I_C$-$H$ patterns are not symmetrical with respect to CI and FI, but are wholly symmetric under CFI as shown by the consistency between $I_C^+(H)$ and $-I_C^-(-H)$ for both cases.
\par
This fact clearly identifies the time-reversal invariance of superconductivity of Sr$_{2}$RuO$_{4}$. To quantitatively evaluate the accuracy of matching, we introduce a common offset field that provides the best consistency between the black and red curves (center of inversion symmetry) by the least squares method. The obtained result of $4.0 \times 10^{-5}$ $H_0$ (0.5 mOe, 0.31 K) and $1.0 \times 10^{-4}$ $H_0$ (8 mOe, 0.32 K) for Junctions X and Y, respectively, are comparable to the residual magnetic field of the present measurement system (4 mOe). In particular, the result of Junction Y is more than three orders of magnitude smaller than the theoretically expected value of 0.24 $H_0$ for the chiral state, and quite consistent with that for helical as shown in Fig. 6. 
\par
%
%
Here we discuss the effect of flux trapping on the present data.
Although both the flux traping and the broken TRS yield similar shift of the main peak in the $I_C-H$ pattern along $H$-axis, these two effects can exactly be discriminated by checking the symmetry of the $I_C-H$ pattern.
Since the trapped flux works as finite external magnetic field, it simply shift the $I_C-H$ with corresponding magnetic field amplitude.
In such case, the CFI symmetry preserved with respect to $H_E$ ($I_C^+(H+H_E)$ and $-I_C^-(-H+H_E)$), here $H_E$ is the external field coupling to the junction.
On the other hand, when the main peak shift is caused by the broken TRS, $I_C-H$ pattern is modified in addition to the peak shift, thus we cannot find  CFI symmetry even if we take acount of finite $H$ shift.
In fact, the red curve in Fig. 6B calcuated for broken TRS SC can never be overlapped to the black curve even if we assume finite $H$ shift.
Therefore, the evaluation of inversion center and the consistency of to the CFI symmetry in the whole $H$ range is quite important to discriminate the two effects.
Based on this consideration, the experimental data of Fig. 5B, C exhibiting the inversion center being almost zero with strict matching to the CFI symmetry in the whole $H$ range lead us to conclude the absence of of flux trapping effect as well as the time reversal invariance of  Sr$_{2}$RuO$_{4}$.
\par
\begin{figure}[t]
  \begin{center}
		\includegraphics[width=1.0\linewidth]{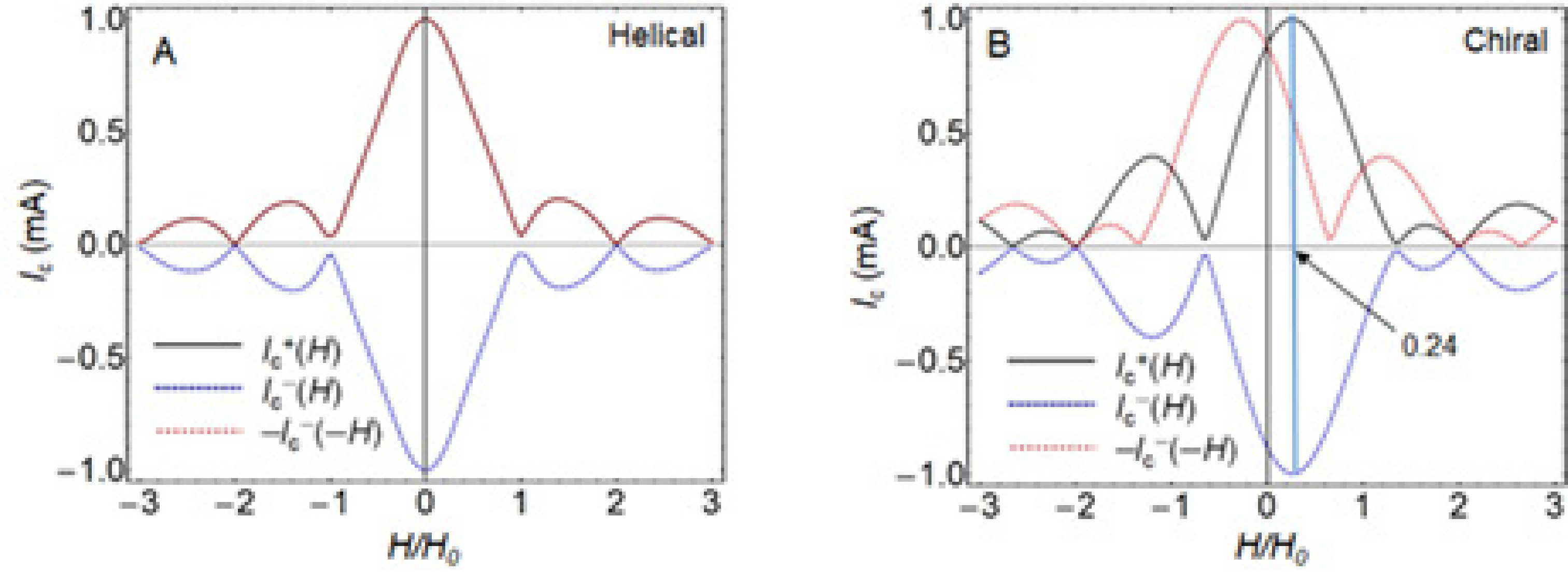}
  \end{center}
  \caption{Theoretically calculated $I_C-H$ symmetry patterns of a corner junction for (A) helical and (B) chiral states based on multiband model developed by Kawai et al. \cite{27}. Assuming corner junction with an apex angle of 2/3$\pi$ with equal current distribution for each wing. The $I_C-H$ pattern for the helical state strictly preserves the current and field inversion (CFI) symmetry, while that for the chiral state breaks the CFI symmetry with a $I_C$ peak shift corresponding to 0.24$H_0$. We expect that experimentally detected inversion center of $I_C-H$ pattern in corner Josephson junction should be comparable to $0.24H_0$ if Sr$_{2}$RuO$_{4}$ is a chiral $p$-wave superconductor assuming that the Josephson current is composed only of the first term.
}
  \label{Fig6}
\end{figure}
\section{Consistency with other experimental results}
As a summary of the experimental results, the basic transport properties are consistent with the triplet SC of Sr$_{2}$RuO$_{4}$. The junction-size dependence of the JJ indicates the multicomponent superconductivity suggested by the presence of the domains~\cite{21,22,30}. The $I_C$-$H$ of the planar and corner junctions indicates time-reversal invariant SC of Sr$_{2}$RuO$_{4}$. Among the candidate pairing states, the helical $p$-wave states do not break the TRS, and simultaneously allow for domain formation owing to the difference in helicity with nearly degenerate energy.
By contrast, TRS $s$-wave,  $d$-wave as well as chiral $p$-wave symmetries cannot account for the multicomponent nature . 
Therefore, we conclude that the present results indicate helical $p$-wave states of Sr$_{2}$RuO$_{4}$.
\par
Although the chiral superconductivity of Sr$_{2}$RuO$_{4}$ has been widely accepted, the helical state is also consistent with many other experiments carried out in recent years. The formation of the topological edge state reproducibly detected by the quasiparticle tunneling effect~\cite{11,12, 50} and the absence of the edge current observed by a variety of real space magnetic field probes~\cite{17,18,19,39} are naturally explained by the helical states. The appearance of the half-integer flux quanta~\cite{51,52} is a consequence of the equal-spin pairing states of Sr$_{2}$RuO$_{4}$, and thus it is compatible with both the helical and chiral states. Furthermore, concerning many other data interpreted in terms of the chiral state in the past, the interpretations are mostly unchanged even for the helical states as long as their measurements are phase insensitive. For example, the multicomponent superconductivity of Sr$_{2}$RuO$_{4}$ detected by magnetization~\cite{51} and in Sr$_{2}$RuO$_{4}$/Ru/Nb~\cite{30} junctions, the super-liner increase in $T_C$ under the uniaxial pressure~\cite{53}, and recent thermal conductivity experiments rejecting the presence of horizontal line nodes~\cite{54} are not in conflict with the helical states.
\par
	On the other hand, the present results apparently conflict with the broken TRS detected in previous experiments~\cite{14,15,37-2}. We discuss how the key experimental results, namely the $\mu$SR, Kerr effect results, and scanning Hall and SQUID results, can be interpreted without conflicting with the present results. We can present two possibilities to resolve this inconsistency. First, let us consider a case where the superconducting state in the bulk is chiral and that at the surface parallel to the $c$-axis is helical. For a second case, we consider that the superconducting state is helical both in the bulk and at the surface. 
\par
In the first case, the interpretation of the $\mu$SR, Kerr effect results are the same as before. The internal field detected by a muon is either induced by a muon or by impurities around the muon. The fact that scanning Hall and SQUID sensors~\cite{17,18,19} do not detect any magnetic field induced by edge current at the sample edge or at the putative chiral domain walls can be explained in theoretical models~\cite{20,40,41,42,43,44,45}; however it requires relatively narrow range of fitting parameters. 
Whereas, the present results urge a novel possibility that a helical state is locally induced at the surface where the junctions are made. The surface helical state may be either intrinsic or induced by the proximity effect from the TRS-preserved $s$-wave superconductivity of Nb. However, since the induced helical state is likely orbitally polarized owing to the coupling to the orbital chirality in the bulk, this should break TRS and exhibits violations in the CFI symmetry invariance.
\par
The second case corresponds to the helical superconductivity in the bulk, as well as the surface, of Sr$_{2}$RuO$_{4}$.  In the helical states, the spin part of the NMR Knight shift is expected to decrease by half for any applied field direction within the ab-plane, which is consistent with recent experimental result~\cite{63}. Whereas one needs to alter the interpretation of the $\mu$SR: a muon or impurities need to induce the chiral state surrounding the non-superconducting core region immediately around the muon or impurities. The field induced at the interface may be probed by the muon. 
In fact, the helical state seems to be extremely sensitive to an applied magnetic field. In our previous JJ experiments, the domain dynamics were induced by increasing magnetic field, exceeding approximately 10 Oe~\cite{22}. Recent scanning Hall probe microscopy reveals the change in the superconducting state with an applied magnetic field of about 25 Oe~\cite{19}. Furthermore, no reduction in the Knight shift was observed either on the $c$-axis or in the in-plane magnetic field directions. This suggests that the $d$-vector can be rotated under the field on the order of 100 Oe~\cite{3,4,8}. All of these results suggest that the superconducting states can be modified readily in the applied magnetic field. Such sensitivity may be one of the clues to account for the inconsistency of previous experiments that supported broken time-reversal symmetry in the presence of muons or local photon irradiation~\cite{14,15}.
Based on these consideration, we conclude that Sr$_{2}$RuO$_{4}$ is a bulk helical $p$-wave SC.
\par
%
%
%
\section{Topological crystalline superconductivity}
With the Cooper-pair spins aligned along the $c$-axis, such as in the helical states discussed here, there is a profound implication concerning the topological nature of the superconductivity of Sr$_{2}$RuO$_{4}$. 
The time-reversal invariance of the helical SC (Fig. 1B and 1C) implies that one of the bulk topological invariants is $Z_2$~\cite{57,58,58-2}. The proper crystalline symmetry of Sr$_{2}$RuO$_{4}$ provides an additional intrinsic topological nature to the helical state~\cite{25}. The crucial point is the mirror reflection symmetry in the crystal structure, as shown in Fig. 1A. 
 It should be first noted that the spin, being an axial vector, obeys the same transformation rule under the mirror reflection as that for the orbital angular momentum of the Cooper pairs. 
While the mirror reflection flips the in-plane components of the electron spins, it avoids the mixture of out-of-plane components. Therefore, the spin-up and spin-down Cooper-pair sectors in Fig. 1B do not mix, and thus these spin sectors behave like two independent fully spin-polarized SCs, as shown in Fig. 1E. Since each spin-polarized SC sector realizes a so-called spinless topological SC, it hosts a single Majorana fermion. 
	Moreover, by just inserting a magnetic flux in the $c$-direction, which maintains the mirror reflection symmetry, each spin sector supports a stable Majorana zero mode. Such a state is referred to as a topological crystalline SC in the analogy of the topological crystalline insulator~\cite{59}. The Majorana modes are stable as long as the mirror reflection is retained. Like an ordinary Majorana zero mode, the mirror-protected Majorana zero modes display non-Abelian statistics; furthermore, the non-Abelian nature can be controlled by slightly breaking the mirror symmetry~\cite{60}. 
	We emphasize that this is in clear contrast to the chiral state with the two-dimensional representation (Fig. 1D). For the chiral state under a tetragonal $D_{4h}$ point-group symmetry, the $d$-vector has to be along the $c$-axis for the following reason. In addition to the mirror reflection, the $d$-vector state should not change under the fourfold rotation ($k_x \rightarrow k_y$, $k_y\rightarrow -k_x$ ; $x \rightarrow y$, $y\rightarrow -x$, $z\rightarrow z$). This requires the spin component to be $z$ rather than $x$ or $y$, and $d = z (k_x \pm i k_y)$. Thus, the Cooper-pair spins are in the ab plane, and opposite-spin Cooper-pair sectors are mixed by the mirror reflection. Note that for the helical states, the same operation maintains the same $d$-vector state as long as the spin components constituting the one-dimensional representations of the order parameter are x and y but not z. Owing to an unavoidable interaction between the two spin sectors, the chiral SC hosts a Dirac fermion rather than Majorana fermions~\cite{25}. An exception is a half-quantum vortex core state where the mirror reflection symmetry is completely broken. Both the chiral and helical states may support a single Majorana edge mode in a half-quantum vortex core~\cite{10,61}. 
	The present result confirms that Sr$_{2}$RuO$_{4}$ is the first example of the topological crystalline SC. In contrast to the chiral SC state, the helical SC state hosts two Majorana modes even in the integer vortex state, facilitating a new approach to the non-Abelian braiding. This is crucially important for future applications of Sr$_{2}$RuO$_{4}$ because the Majorana zero mode is the essential ingredient in realizing a fault-tolerant topological quantum computation~\cite{62}.
\par
%
%
\section{Summaries}
To identify the time-reversal symmetry of Sr$_{2}$RuO$_{4}$, we propose a current-field inversion (CFI) test that examines the invariance of the Josephson critical current under the inversion of both the current and magnetic fields. 
By applying this method to the Josephson junctions between Sr$_{2}$RuO$_{4}$ and Nb, we conclude that the superconductivity of Sr$_{2}$RuO$_{4}$ is verified as time-reversal invariant and multicomponental as indicated by the CFI symmetry of $I_C-H$ patterns and by the presence of dynamical domains. Among the table of the Sr$_{2}$RuO$_{4}$ candidates, we believe helical $p$-wave superconductor is most reasonable. Although the energy level of the each helical states are not degenerated in general, the difference is quite small in Sr$_{2}$RuO$_{4}$ so that two of them could coexist and form the domain structure.  However, we may need to discuss more complex possibilities, such as singlet-triplet mixing, as well as the stability of the helical domains, in future works. We discuss how this new finding is compatible with previous claims of TRS breaking. The present conclusion identifies Sr$_{2}$RuO$_{4}$ as the first discovery of a topological crystalline superconductor that can host two Majorana edge modes at the surface protected by crystalline mirror symmetry. 
\par
\section{Acknowledgments}
We acknowledge S. Kittaka and T. Sumi for the crystal growth; Y. Asano for valuable discussions; K. Tsumura for drawing figures. This work was supported by JSPS KAKENHI (nos. JP15H05851, JP15H05852, JP15H05853, JP15H05855, JP15K21717, JP17H02922, and 18H01243), the Oxide Superspin (OSS) Core-to-Core Program, as well as CREST, JST (no. JPMJCR16F2).

%
%

%


\bibliography{SRO1.bib}

\begin{thebibliography}{63}%
\makeatletter
\providecommand \@ifxundefined [1]{%
 \@ifx{#1\undefined}
}%
\providecommand \@ifnum [1]{%
 \ifnum #1\expandafter \@firstoftwo
 \else \expandafter \@secondoftwo
 \fi
}%
\providecommand \@ifx [1]{%
 \ifx #1\expandafter \@firstoftwo
 \else \expandafter \@secondoftwo
 \fi
}%
\providecommand \natexlab [1]{#1}%
\providecommand \enquote  [1]{``#1''}%
\providecommand \bibnamefont  [1]{#1}%
\providecommand \bibfnamefont [1]{#1}%
\providecommand \citenamefont [1]{#1}%
\providecommand \href@noop [0]{\@secondoftwo}%
\providecommand \href [0]{\begingroup \@sanitize@url \@href}%
\providecommand \@href[1]{\@@startlink{#1}\@@href}%
\providecommand \@@href[1]{\endgroup#1\@@endlink}%
\providecommand \@sanitize@url [0]{\catcode `\\12\catcode `\$12\catcode
  `\&12\catcode `\#12\catcode `\^12\catcode `\_12\catcode `\%12\relax}%
\providecommand \@@startlink[1]{}%
\providecommand \@@endlink[0]{}%
\providecommand \url  [0]{\begingroup\@sanitize@url \@url }%
\providecommand \@url [1]{\endgroup\@href {#1}{\urlprefix }}%
\providecommand \urlprefix  [0]{URL }%
\providecommand \Eprint [0]{\href }%
\providecommand \doibase [0]{http://dx.doi.org/}%
\providecommand \selectlanguage [0]{\@gobble}%
\providecommand \bibinfo  [0]{\@secondoftwo}%
\providecommand \bibfield  [0]{\@secondoftwo}%
\providecommand \translation [1]{[#1]}%
\providecommand \BibitemOpen [0]{}%
\providecommand \bibitemStop [0]{}%
\providecommand \bibitemNoStop [0]{.\EOS\space}%
\providecommand \EOS [0]{\spacefactor3000\relax}%
\providecommand \BibitemShut  [1]{\csname bibitem#1\endcsname}%
\let\auto@bib@innerbib\@empty
\bibitem [{\citenamefont {Bardeen}\ \emph {et~al.}(1957)\citenamefont
  {Bardeen}, \citenamefont {Cooper},\ and\ \citenamefont {Schrieffer}}]{1}%
  \BibitemOpen
  \bibfield  {author} {\bibinfo {author} {\bibfnamefont {J.}~\bibnamefont
  {Bardeen}}, \bibinfo {author} {\bibfnamefont {L.~N.}\ \bibnamefont {Cooper}},
  \ and\ \bibinfo {author} {\bibfnamefont {J.~R.}\ \bibnamefont {Schrieffer}},\
  }\href@noop {} {\bibfield  {journal} {\bibinfo  {journal} {Phys. Rev.}\
  }\textbf {\bibinfo {volume} {108}},\ \bibinfo {pages} {1175} (\bibinfo {year}
  {1957})}\BibitemShut {NoStop}%
\bibitem [{\citenamefont {Maeno}\ \emph {et~al.}(1994)\citenamefont {Maeno},
  \citenamefont {Hashimoto}, \citenamefont {Yoshida}, \citenamefont
  {Nishizaki}, \citenamefont {Fujita}, \citenamefont {Bednorz},\ and\
  \citenamefont {Lichtenberg}}]{2}%
  \BibitemOpen
  \bibfield  {author} {\bibinfo {author} {\bibfnamefont {Y.}~\bibnamefont
  {Maeno}}, \bibinfo {author} {\bibfnamefont {H.}~\bibnamefont {Hashimoto}},
  \bibinfo {author} {\bibfnamefont {K.}~\bibnamefont {Yoshida}}, \bibinfo
  {author} {\bibfnamefont {S.}~\bibnamefont {Nishizaki}}, \bibinfo {author}
  {\bibfnamefont {T.}~\bibnamefont {Fujita}}, \bibinfo {author} {\bibfnamefont
  {J.~G.}\ \bibnamefont {Bednorz}}, \ and\ \bibinfo {author} {\bibfnamefont
  {F.}~\bibnamefont {Lichtenberg}},\ }\href@noop {} {\bibfield  {journal}
  {\bibinfo  {journal} {Nature}\ }\textbf {\bibinfo {volume} {372}},\ \bibinfo
  {pages} {352} (\bibinfo {year} {1994})}\BibitemShut {NoStop}%
\bibitem [{\citenamefont {Mackenzie}\ and\ \citenamefont {Maeno}(2003)}]{3}%
  \BibitemOpen
  \bibfield  {author} {\bibinfo {author} {\bibfnamefont {A.~P.}\ \bibnamefont
  {Mackenzie}}\ and\ \bibinfo {author} {\bibfnamefont {Y.}~\bibnamefont
  {Maeno}},\ }\href@noop {} {\bibfield  {journal} {\bibinfo  {journal} {Rev.
  Mod. Phys.}\ }\textbf {\bibinfo {volume} {75}},\ \bibinfo {pages} {657}
  (\bibinfo {year} {2003})}\BibitemShut {NoStop}%
\bibitem [{\citenamefont {Maeno}\ \emph {et~al.}(2012)\citenamefont {Maeno},
  \citenamefont {Kittaka}, \citenamefont {Nomura}, \citenamefont {Yonezawa},\
  and\ \citenamefont {Ishida}}]{4}%
  \BibitemOpen
  \bibfield  {author} {\bibinfo {author} {\bibfnamefont {Y.}~\bibnamefont
  {Maeno}}, \bibinfo {author} {\bibfnamefont {S.}~\bibnamefont {Kittaka}},
  \bibinfo {author} {\bibfnamefont {T.}~\bibnamefont {Nomura}}, \bibinfo
  {author} {\bibfnamefont {S.}~\bibnamefont {Yonezawa}}, \ and\ \bibinfo
  {author} {\bibfnamefont {K.}~\bibnamefont {Ishida}},\ }\href@noop {}
  {\bibfield  {journal} {\bibinfo  {journal} {J. Phys. Soc. Jpn.}\ }\textbf
  {\bibinfo {volume} {81}},\ \bibinfo {pages} {011009} (\bibinfo {year}
  {2012})}\BibitemShut {NoStop}%
\bibitem [{\citenamefont {Kallin}(2012)}]{5}%
  \BibitemOpen
  \bibfield  {author} {\bibinfo {author} {\bibfnamefont {C.}~\bibnamefont
  {Kallin}},\ }\href@noop {} {\bibfield  {journal} {\bibinfo  {journal} {Rep.
  Prog. Phys.}\ }\textbf {\bibinfo {volume} {75}},\ \bibinfo {pages} {042501}
  (\bibinfo {year} {2012})}\BibitemShut {NoStop}%
\bibitem [{\citenamefont {Liu}\ and\ \citenamefont {Mao}(2015)}]{6}%
  \BibitemOpen
  \bibfield  {author} {\bibinfo {author} {\bibfnamefont {Y.}~\bibnamefont
  {Liu}}\ and\ \bibinfo {author} {\bibfnamefont {Z.-Q.}\ \bibnamefont {Mao}},\
  }\href@noop {} {\bibfield  {journal} {\bibinfo  {journal} {Phys. C-Supercond.
  Appl.}\ }\textbf {\bibinfo {volume} {514}},\ \bibinfo {pages} {339} (\bibinfo
  {year} {2015})}\BibitemShut {NoStop}%
\bibitem [{\citenamefont {Mackenzie}\ \emph {et~al.}(2017)\citenamefont
  {Mackenzie}, \citenamefont {Scaffidi}, \citenamefont {Hicks},\ and\
  \citenamefont {Maeno}}]{7}%
  \BibitemOpen
  \bibfield  {author} {\bibinfo {author} {\bibfnamefont {A.~P.}\ \bibnamefont
  {Mackenzie}}, \bibinfo {author} {\bibfnamefont {T.}~\bibnamefont {Scaffidi}},
  \bibinfo {author} {\bibfnamefont {C.~W.}\ \bibnamefont {Hicks}}, \ and\
  \bibinfo {author} {\bibfnamefont {Y.}~\bibnamefont {Maeno}},\ }\href@noop {}
  {\bibfield  {journal} {\bibinfo  {journal} {NPJ Quantum Materials}\ }\textbf
  {\bibinfo {volume} {2}},\ \bibinfo {pages} {1} (\bibinfo {year}
  {2017})}\BibitemShut {NoStop}%
\bibitem [{\citenamefont {Murakawa}\ \emph {et~al.}(2004)\citenamefont
  {Murakawa}, \citenamefont {Ishida}, \citenamefont {Kitagawa}, \citenamefont
  {Mao},\ and\ \citenamefont {Maeno}}]{8}%
  \BibitemOpen
  \bibfield  {author} {\bibinfo {author} {\bibfnamefont {H.}~\bibnamefont
  {Murakawa}}, \bibinfo {author} {\bibfnamefont {K.}~\bibnamefont {Ishida}},
  \bibinfo {author} {\bibfnamefont {K.}~\bibnamefont {Kitagawa}}, \bibinfo
  {author} {\bibfnamefont {Z.~Q.}\ \bibnamefont {Mao}}, \ and\ \bibinfo
  {author} {\bibfnamefont {Y.}~\bibnamefont {Maeno}},\ }\href@noop {}
  {\bibfield  {journal} {\bibinfo  {journal} {Phys. Rev. Lett.}\ }\textbf
  {\bibinfo {volume} {93}},\ \bibinfo {pages} {167004} (\bibinfo {year}
  {2004})}\BibitemShut {NoStop}%
\bibitem [{\citenamefont {Nelson}\ \emph {et~al.}(2004)\citenamefont {Nelson},
  \citenamefont {Mao}, \citenamefont {Maeno},\ and\ \citenamefont {Liu}}]{9}%
  \BibitemOpen
  \bibfield  {author} {\bibinfo {author} {\bibfnamefont {K.~D.}\ \bibnamefont
  {Nelson}}, \bibinfo {author} {\bibfnamefont {Z.~Q.}\ \bibnamefont {Mao}},
  \bibinfo {author} {\bibfnamefont {Y.}~\bibnamefont {Maeno}}, \ and\ \bibinfo
  {author} {\bibfnamefont {Y.}~\bibnamefont {Liu}},\ }\href@noop {} {\bibfield
  {journal} {\bibinfo  {journal} {Science}\ }\textbf {\bibinfo {volume}
  {306}},\ \bibinfo {pages} {1151} (\bibinfo {year} {2004})}\BibitemShut
  {NoStop}%
\bibitem [{\citenamefont {Read}\ and\ \citenamefont {Green}(2000)}]{10}%
  \BibitemOpen
  \bibfield  {author} {\bibinfo {author} {\bibfnamefont {N.}~\bibnamefont
  {Read}}\ and\ \bibinfo {author} {\bibfnamefont {D.}~\bibnamefont {Green}},\
  }\href@noop {} {\bibfield  {journal} {\bibinfo  {journal} {Phys. Rev. B}\
  }\textbf {\bibinfo {volume} {61}},\ \bibinfo {pages} {10267} (\bibinfo {year}
  {2000})}\BibitemShut {NoStop}%
\bibitem [{\citenamefont {Laube}\ \emph {et~al.}(2000)\citenamefont {Laube},
  \citenamefont {Goll}, \citenamefont {v.~Löhneysen}, \citenamefont
  {Fogelström},\ and\ \citenamefont {Lichtenberg}}]{11}%
  \BibitemOpen
  \bibfield  {author} {\bibinfo {author} {\bibfnamefont {F.}~\bibnamefont
  {Laube}}, \bibinfo {author} {\bibfnamefont {G.}~\bibnamefont {Goll}},
  \bibinfo {author} {\bibfnamefont {H.}~\bibnamefont {v.~Löhneysen}}, \bibinfo
  {author} {\bibfnamefont {M.}~\bibnamefont {Fogelström}}, \ and\ \bibinfo
  {author} {\bibfnamefont {F.}~\bibnamefont {Lichtenberg}},\ }\href@noop {}
  {\bibfield  {journal} {\bibinfo  {journal} {Phys. Rev. Lett.}\ }\textbf
  {\bibinfo {volume} {84}},\ \bibinfo {pages} {1595} (\bibinfo {year}
  {2000})}\BibitemShut {NoStop}%
\bibitem [{\citenamefont {Kashiwaya}\ \emph {et~al.}(2011)\citenamefont
  {Kashiwaya}, \citenamefont {Kashiwaya}, \citenamefont {Kambara},
  \citenamefont {Furuta}, \citenamefont {Yaguchi}, \citenamefont {Tanaka},\
  and\ \citenamefont {Maeno}}]{12}%
  \BibitemOpen
  \bibfield  {author} {\bibinfo {author} {\bibfnamefont {S.}~\bibnamefont
  {Kashiwaya}}, \bibinfo {author} {\bibfnamefont {H.}~\bibnamefont
  {Kashiwaya}}, \bibinfo {author} {\bibfnamefont {H.}~\bibnamefont {Kambara}},
  \bibinfo {author} {\bibfnamefont {T.}~\bibnamefont {Furuta}}, \bibinfo
  {author} {\bibfnamefont {H.}~\bibnamefont {Yaguchi}}, \bibinfo {author}
  {\bibfnamefont {Y.}~\bibnamefont {Tanaka}}, \ and\ \bibinfo {author}
  {\bibfnamefont {Y.}~\bibnamefont {Maeno}},\ }\href@noop {} {\bibfield
  {journal} {\bibinfo  {journal} {Phys. Rev. Lett.}\ }\textbf {\bibinfo
  {volume} {107}},\ \bibinfo {pages} {077003} (\bibinfo {year}
  {2011})}\BibitemShut {NoStop}%
\bibitem [{\citenamefont {Wang}\ \emph {et~al.}(2015)\citenamefont {Wang},
  \citenamefont {Lou}, \citenamefont {Luo}, \citenamefont {Wei}, \citenamefont
  {Liu}, \citenamefont {Ortmann},\ and\ \citenamefont {Mao}}]{50}%
  \BibitemOpen
  \bibfield  {author} {\bibinfo {author} {\bibfnamefont {H.}~\bibnamefont
  {Wang}}, \bibinfo {author} {\bibfnamefont {W.}~\bibnamefont {Lou}}, \bibinfo
  {author} {\bibfnamefont {J.}~\bibnamefont {Luo}}, \bibinfo {author}
  {\bibfnamefont {J.}~\bibnamefont {Wei}}, \bibinfo {author} {\bibfnamefont
  {Y.}~\bibnamefont {Liu}}, \bibinfo {author} {\bibfnamefont {J.~E.}\
  \bibnamefont {Ortmann}}, \ and\ \bibinfo {author} {\bibfnamefont {Z.~Q.}\
  \bibnamefont {Mao}},\ }\href@noop {} {\bibfield  {journal} {\bibinfo
  {journal} {Phys. Rev. B}\ }\textbf {\bibinfo {volume} {91}},\ \bibinfo
  {pages} {184514} (\bibinfo {year} {2015})}\BibitemShut {NoStop}%
\bibitem [{\citenamefont {Scaffidi}\ \emph {et~al.}(2014)\citenamefont
  {Scaffidi}, \citenamefont {Romers},\ and\ \citenamefont {Simon}}]{13}%
  \BibitemOpen
  \bibfield  {author} {\bibinfo {author} {\bibfnamefont {T.}~\bibnamefont
  {Scaffidi}}, \bibinfo {author} {\bibfnamefont {J.~C.}\ \bibnamefont
  {Romers}}, \ and\ \bibinfo {author} {\bibfnamefont {S.~H.}\ \bibnamefont
  {Simon}},\ }\href@noop {} {\bibfield  {journal} {\bibinfo  {journal} {Phys.
  Rev. B}\ }\textbf {\bibinfo {volume} {89}},\ \bibinfo {pages} {220510(R)}
  (\bibinfo {year} {2014})}\BibitemShut {NoStop}%
\bibitem [{\citenamefont {Luke}\ \emph {et~al.}(1999)\citenamefont {Luke},
  \citenamefont {Fudamoto}, \citenamefont {Kojima}, \citenamefont {Larkin},
  \citenamefont {Merrin}, \citenamefont {Nachumi}, \citenamefont {Uemura},
  \citenamefont {Maeno}, \citenamefont {Mao}, \citenamefont {Mori},
  \citenamefont {Nakamura},\ and\ \citenamefont {Sigrist}}]{14}%
  \BibitemOpen
  \bibfield  {author} {\bibinfo {author} {\bibfnamefont {G.~M.}\ \bibnamefont
  {Luke}}, \bibinfo {author} {\bibfnamefont {Y.}~\bibnamefont {Fudamoto}},
  \bibinfo {author} {\bibfnamefont {K.~M.}\ \bibnamefont {Kojima}}, \bibinfo
  {author} {\bibfnamefont {M.~I.}\ \bibnamefont {Larkin}}, \bibinfo {author}
  {\bibfnamefont {J.}~\bibnamefont {Merrin}}, \bibinfo {author} {\bibfnamefont
  {B.}~\bibnamefont {Nachumi}}, \bibinfo {author} {\bibfnamefont {Y.~J.}\
  \bibnamefont {Uemura}}, \bibinfo {author} {\bibfnamefont {Y.}~\bibnamefont
  {Maeno}}, \bibinfo {author} {\bibfnamefont {Z.~Q.}\ \bibnamefont {Mao}},
  \bibinfo {author} {\bibfnamefont {Y.}~\bibnamefont {Mori}}, \bibinfo {author}
  {\bibfnamefont {H.}~\bibnamefont {Nakamura}}, \ and\ \bibinfo {author}
  {\bibfnamefont {M.}~\bibnamefont {Sigrist}},\ }\href@noop {} {\bibfield
  {journal} {\bibinfo  {journal} {Nature}\ }\textbf {\bibinfo {volume} {394}},\
  \bibinfo {pages} {558} (\bibinfo {year} {1999})}\BibitemShut {NoStop}%
\bibitem [{\citenamefont {Xia}\ \emph {et~al.}(2006)\citenamefont {Xia},
  \citenamefont {Maeno}, \citenamefont {Beyersdorf}, \citenamefont {Fejer},\
  and\ \citenamefont {Kapitulnik}}]{15}%
  \BibitemOpen
  \bibfield  {author} {\bibinfo {author} {\bibfnamefont {J.}~\bibnamefont
  {Xia}}, \bibinfo {author} {\bibfnamefont {Y.}~\bibnamefont {Maeno}}, \bibinfo
  {author} {\bibfnamefont {P.~T.}\ \bibnamefont {Beyersdorf}}, \bibinfo
  {author} {\bibfnamefont {M.~M.}\ \bibnamefont {Fejer}}, \ and\ \bibinfo
  {author} {\bibfnamefont {A.}~\bibnamefont {Kapitulnik}},\ }\href@noop {}
  {\bibfield  {journal} {\bibinfo  {journal} {Phys. Rev. Lett.}\ }\textbf
  {\bibinfo {volume} {97}},\ \bibinfo {pages} {167002} (\bibinfo {year}
  {2006})}\BibitemShut {NoStop}%
\bibitem [{\citenamefont {Matsumoto}\ and\ \citenamefont
  {Sigrist}(1999{\natexlab{a}})}]{16}%
  \BibitemOpen
  \bibfield  {author} {\bibinfo {author} {\bibfnamefont {M.}~\bibnamefont
  {Matsumoto}}\ and\ \bibinfo {author} {\bibfnamefont {M.}~\bibnamefont
  {Sigrist}},\ }\href@noop {} {\bibfield  {journal} {\bibinfo  {journal} {J.
  Phys. Soc. Jpn.}\ }\textbf {\bibinfo {volume} {68}},\ \bibinfo {pages} {994}
  (\bibinfo {year} {1999}{\natexlab{a}})}\BibitemShut {NoStop}%
\bibitem [{\citenamefont {Matsumoto}\ and\ \citenamefont
  {Sigrist}(1999{\natexlab{b}})}]{16-2}%
  \BibitemOpen
  \bibfield  {author} {\bibinfo {author} {\bibfnamefont {M.}~\bibnamefont
  {Matsumoto}}\ and\ \bibinfo {author} {\bibfnamefont {M.}~\bibnamefont
  {Sigrist}},\ }\href@noop {} {\bibfield  {journal} {\bibinfo  {journal} {J.
  Phys. Soc. Jpn.}\ }\textbf {\bibinfo {volume} {68}},\ \bibinfo {pages} {3120}
  (\bibinfo {year} {1999}{\natexlab{b}})}\BibitemShut {NoStop}%
\bibitem [{\citenamefont {Kirtley}\ \emph {et~al.}(2007)\citenamefont
  {Kirtley}, \citenamefont {Kallin}, \citenamefont {Hicks}, \citenamefont
  {Kim}, \citenamefont {Liu}, \citenamefont {Moler}, \citenamefont {Maeno},\
  and\ \citenamefont {Nelson}}]{17}%
  \BibitemOpen
  \bibfield  {author} {\bibinfo {author} {\bibfnamefont {J.~R.}\ \bibnamefont
  {Kirtley}}, \bibinfo {author} {\bibfnamefont {C.}~\bibnamefont {Kallin}},
  \bibinfo {author} {\bibfnamefont {C.~W.}\ \bibnamefont {Hicks}}, \bibinfo
  {author} {\bibfnamefont {E.-A.}\ \bibnamefont {Kim}}, \bibinfo {author}
  {\bibfnamefont {Y.}~\bibnamefont {Liu}}, \bibinfo {author} {\bibfnamefont
  {K.~A.}\ \bibnamefont {Moler}}, \bibinfo {author} {\bibfnamefont
  {Y.}~\bibnamefont {Maeno}}, \ and\ \bibinfo {author} {\bibfnamefont {K.~D.}\
  \bibnamefont {Nelson}},\ }\href@noop {} {\bibfield  {journal} {\bibinfo
  {journal} {Phys. Rev. B}\ }\textbf {\bibinfo {volume} {76}},\ \bibinfo
  {pages} {014526} (\bibinfo {year} {2007})}\BibitemShut {NoStop}%
\bibitem [{\citenamefont {Hicks}\ \emph {et~al.}(2010)\citenamefont {Hicks},
  \citenamefont {Kirtley}, \citenamefont {Lippman}, \citenamefont {Koshnick},
  \citenamefont {Huber}, \citenamefont {Maeno}, \citenamefont {Yuhasz},
  \citenamefont {Maple},\ and\ \citenamefont {Moler}}]{18}%
  \BibitemOpen
  \bibfield  {author} {\bibinfo {author} {\bibfnamefont {C.~W.}\ \bibnamefont
  {Hicks}}, \bibinfo {author} {\bibfnamefont {J.~R.}\ \bibnamefont {Kirtley}},
  \bibinfo {author} {\bibfnamefont {T.~M.}\ \bibnamefont {Lippman}}, \bibinfo
  {author} {\bibfnamefont {N.~C.}\ \bibnamefont {Koshnick}}, \bibinfo {author}
  {\bibfnamefont {M.~E.}\ \bibnamefont {Huber}}, \bibinfo {author}
  {\bibfnamefont {Y.}~\bibnamefont {Maeno}}, \bibinfo {author} {\bibfnamefont
  {W.~M.}\ \bibnamefont {Yuhasz}}, \bibinfo {author} {\bibfnamefont {M.~B.}\
  \bibnamefont {Maple}}, \ and\ \bibinfo {author} {\bibfnamefont {K.~A.}\
  \bibnamefont {Moler}},\ }\href@noop {} {\bibfield  {journal} {\bibinfo
  {journal} {Phys. Rev. B}\ }\textbf {\bibinfo {volume} {81}},\ \bibinfo
  {pages} {214501} (\bibinfo {year} {2010})}\BibitemShut {NoStop}%
\bibitem [{\citenamefont {Curran}\ \emph {et~al.}(2014)\citenamefont {Curran},
  \citenamefont {Bending}, \citenamefont {Desoky}, \citenamefont {Gibbs},
  \citenamefont {Lee},\ and\ \citenamefont {Mackenzie}}]{19}%
  \BibitemOpen
  \bibfield  {author} {\bibinfo {author} {\bibfnamefont {P.~J.}\ \bibnamefont
  {Curran}}, \bibinfo {author} {\bibfnamefont {S.~J.}\ \bibnamefont {Bending}},
  \bibinfo {author} {\bibfnamefont {W.~M.}\ \bibnamefont {Desoky}}, \bibinfo
  {author} {\bibfnamefont {A.~S.}\ \bibnamefont {Gibbs}}, \bibinfo {author}
  {\bibfnamefont {S.~L.}\ \bibnamefont {Lee}}, \ and\ \bibinfo {author}
  {\bibfnamefont {A.~P.}\ \bibnamefont {Mackenzie}},\ }\href@noop {} {\bibfield
   {journal} {\bibinfo  {journal} {Phys. Rev. B}\ }\textbf {\bibinfo {volume}
  {89}},\ \bibinfo {pages} {144504} (\bibinfo {year} {2014})}\BibitemShut
  {NoStop}%
\bibitem [{\citenamefont {Huang}\ \emph {et~al.}(2015)\citenamefont {Huang},
  \citenamefont {Lederer}, \citenamefont {Taylor},\ and\ \citenamefont
  {Kallin}}]{20}%
  \BibitemOpen
  \bibfield  {author} {\bibinfo {author} {\bibfnamefont {W.}~\bibnamefont
  {Huang}}, \bibinfo {author} {\bibfnamefont {S.}~\bibnamefont {Lederer}},
  \bibinfo {author} {\bibfnamefont {E.}~\bibnamefont {Taylor}}, \ and\ \bibinfo
  {author} {\bibfnamefont {C.}~\bibnamefont {Kallin}},\ }\href@noop {}
  {\bibfield  {journal} {\bibinfo  {journal} {Phys. Rev. B}\ }\textbf {\bibinfo
  {volume} {91}},\ \bibinfo {pages} {094507} (\bibinfo {year}
  {2015})}\BibitemShut {NoStop}%
\bibitem [{\citenamefont {Ashby}\ and\ \citenamefont {Kallin}(2009)}]{40}%
  \BibitemOpen
  \bibfield  {author} {\bibinfo {author} {\bibfnamefont {P.~E.~C.}\
  \bibnamefont {Ashby}}\ and\ \bibinfo {author} {\bibfnamefont
  {C.}~\bibnamefont {Kallin}},\ }\href@noop {} {\bibfield  {journal} {\bibinfo
  {journal} {Phys. Rev. B}\ }\textbf {\bibinfo {volume} {79}},\ \bibinfo
  {pages} {224509} (\bibinfo {year} {2009})}\BibitemShut {NoStop}%
\bibitem [{\citenamefont {Raghu}\ \emph {et~al.}(2010)\citenamefont {Raghu},
  \citenamefont {Kapitulnik},\ and\ \citenamefont {Kivelson}}]{41}%
  \BibitemOpen
  \bibfield  {author} {\bibinfo {author} {\bibfnamefont {S.}~\bibnamefont
  {Raghu}}, \bibinfo {author} {\bibfnamefont {A.}~\bibnamefont {Kapitulnik}}, \
  and\ \bibinfo {author} {\bibfnamefont {S.~A.}\ \bibnamefont {Kivelson}},\
  }\href@noop {} {\bibfield  {journal} {\bibinfo  {journal} {Phys. Rev. Lett.}\
  }\textbf {\bibinfo {volume} {105}},\ \bibinfo {pages} {136401} (\bibinfo
  {year} {2010})}\BibitemShut {NoStop}%
\bibitem [{\citenamefont {Imai}\ \emph {et~al.}(2012)\citenamefont {Imai},
  \citenamefont {Wakabayashi},\ and\ \citenamefont {Sigrist}}]{42}%
  \BibitemOpen
  \bibfield  {author} {\bibinfo {author} {\bibfnamefont {Y.}~\bibnamefont
  {Imai}}, \bibinfo {author} {\bibfnamefont {K.}~\bibnamefont {Wakabayashi}}, \
  and\ \bibinfo {author} {\bibfnamefont {M.}~\bibnamefont {Sigrist}},\
  }\href@noop {} {\bibfield  {journal} {\bibinfo  {journal} {Phys. Rev. B}\
  }\textbf {\bibinfo {volume} {85}},\ \bibinfo {pages} {174532} (\bibinfo
  {year} {2012})}\BibitemShut {NoStop}%
\bibitem [{\citenamefont {Huang}\ \emph {et~al.}(2014)\citenamefont {Huang},
  \citenamefont {Taylor},\ and\ \citenamefont {Kallin}}]{43}%
  \BibitemOpen
  \bibfield  {author} {\bibinfo {author} {\bibfnamefont {W.}~\bibnamefont
  {Huang}}, \bibinfo {author} {\bibfnamefont {E.}~\bibnamefont {Taylor}}, \
  and\ \bibinfo {author} {\bibfnamefont {C.}~\bibnamefont {Kallin}},\
  }\href@noop {} {\bibfield  {journal} {\bibinfo  {journal} {Phys. Rev. B}\
  }\textbf {\bibinfo {volume} {90}},\ \bibinfo {pages} {224519} (\bibinfo
  {year} {2014})}\BibitemShut {NoStop}%
\bibitem [{\citenamefont {Lederer}\ \emph {et~al.}(2014)\citenamefont
  {Lederer}, \citenamefont {Huang}, \citenamefont {Taylor}, \citenamefont
  {Raghu},\ and\ \citenamefont {Kallin}}]{44}%
  \BibitemOpen
  \bibfield  {author} {\bibinfo {author} {\bibfnamefont {S.}~\bibnamefont
  {Lederer}}, \bibinfo {author} {\bibfnamefont {W.}~\bibnamefont {Huang}},
  \bibinfo {author} {\bibfnamefont {E.}~\bibnamefont {Taylor}}, \bibinfo
  {author} {\bibfnamefont {S.}~\bibnamefont {Raghu}}, \ and\ \bibinfo {author}
  {\bibfnamefont {C.}~\bibnamefont {Kallin}},\ }\href@noop {} {\bibfield
  {journal} {\bibinfo  {journal} {Phys. Rev. B}\ }\textbf {\bibinfo {volume}
  {90}},\ \bibinfo {pages} {134521} (\bibinfo {year} {2014})}\BibitemShut
  {NoStop}%
\bibitem [{\citenamefont {Scaffidi}\ and\ \citenamefont {Simon}(2015)}]{45}%
  \BibitemOpen
  \bibfield  {author} {\bibinfo {author} {\bibfnamefont {T.}~\bibnamefont
  {Scaffidi}}\ and\ \bibinfo {author} {\bibfnamefont {S.~H.}\ \bibnamefont
  {Simon}},\ }\href@noop {} {\bibfield  {journal} {\bibinfo  {journal} {Phys.
  Rev. Lett.}\ }\textbf {\bibinfo {volume} {115}},\ \bibinfo {pages} {087003}
  (\bibinfo {year} {2015})}\BibitemShut {NoStop}%
\bibitem [{\citenamefont {Kidwingira}\ \emph {et~al.}(2006)\citenamefont
  {Kidwingira}, \citenamefont {Strand}, \citenamefont {Harlingen},\ and\
  \citenamefont {Maeno}}]{21}%
  \BibitemOpen
  \bibfield  {author} {\bibinfo {author} {\bibfnamefont {F.}~\bibnamefont
  {Kidwingira}}, \bibinfo {author} {\bibfnamefont {J.~D.}\ \bibnamefont
  {Strand}}, \bibinfo {author} {\bibfnamefont {D.~J.~V.}\ \bibnamefont
  {Harlingen}}, \ and\ \bibinfo {author} {\bibfnamefont {Y.}~\bibnamefont
  {Maeno}},\ }\href@noop {} {\bibfield  {journal} {\bibinfo  {journal}
  {Science}\ }\textbf {\bibinfo {volume} {314}},\ \bibinfo {pages} {1267}
  (\bibinfo {year} {2006})}\BibitemShut {NoStop}%
\bibitem [{\citenamefont {Saitoh}\ \emph {et~al.}(2015)\citenamefont {Saitoh},
  \citenamefont {Kashiwaya}, \citenamefont {Kashiwaya}, \citenamefont
  {Mawatari}, \citenamefont {Asano}, \citenamefont {Tanaka},\ and\
  \citenamefont {Maeno}}]{22}%
  \BibitemOpen
  \bibfield  {author} {\bibinfo {author} {\bibfnamefont {K.}~\bibnamefont
  {Saitoh}}, \bibinfo {author} {\bibfnamefont {S.}~\bibnamefont {Kashiwaya}},
  \bibinfo {author} {\bibfnamefont {H.}~\bibnamefont {Kashiwaya}}, \bibinfo
  {author} {\bibfnamefont {Y.}~\bibnamefont {Mawatari}}, \bibinfo {author}
  {\bibfnamefont {Y.}~\bibnamefont {Asano}}, \bibinfo {author} {\bibfnamefont
  {Y.}~\bibnamefont {Tanaka}}, \ and\ \bibinfo {author} {\bibfnamefont
  {Y.}~\bibnamefont {Maeno}},\ }\href@noop {} {\bibfield  {journal} {\bibinfo
  {journal} {Phys. Rev. B}\ }\textbf {\bibinfo {volume} {92}},\ \bibinfo
  {pages} {100504(R)} (\bibinfo {year} {2015})}\BibitemShut {NoStop}%
\bibitem [{\citenamefont {Geshkenbein}\ \emph {et~al.}(1987)\citenamefont
  {Geshkenbein}, \citenamefont {Larkin},\ and\ \citenamefont {Barone}}]{23}%
  \BibitemOpen
  \bibfield  {author} {\bibinfo {author} {\bibfnamefont {V.~B.}\ \bibnamefont
  {Geshkenbein}}, \bibinfo {author} {\bibfnamefont {A.~I.}\ \bibnamefont
  {Larkin}}, \ and\ \bibinfo {author} {\bibfnamefont {A.}~\bibnamefont
  {Barone}},\ }\href@noop {} {\bibfield  {journal} {\bibinfo  {journal} {Phys.
  Rev. B}\ }\textbf {\bibinfo {volume} {36}},\ \bibinfo {pages} {235} (\bibinfo
  {year} {1987})}\BibitemShut {NoStop}%
\bibitem [{\citenamefont {Harlingen}(1995)}]{24}%
  \BibitemOpen
  \bibfield  {author} {\bibinfo {author} {\bibfnamefont {D.~J.~V.}\
  \bibnamefont {Harlingen}},\ }\href@noop {} {\bibfield  {journal} {\bibinfo
  {journal} {Rev. Mod. Phys.}\ }\textbf {\bibinfo {volume} {67}},\ \bibinfo
  {pages} {515} (\bibinfo {year} {1995})}\BibitemShut {NoStop}%
\bibitem [{\citenamefont {Beasley}\ \emph {et~al.}(1994)\citenamefont
  {Beasley}, \citenamefont {Lew},\ and\ \citenamefont {Laughlin}}]{64}%
  \BibitemOpen
  \bibfield  {author} {\bibinfo {author} {\bibfnamefont {M.~R.}\ \bibnamefont
  {Beasley}}, \bibinfo {author} {\bibfnamefont {D.}~\bibnamefont {Lew}}, \ and\
  \bibinfo {author} {\bibfnamefont {R.~B.}\ \bibnamefont {Laughlin}},\
  }\href@noop {} {\bibfield  {journal} {\bibinfo  {journal} {Phys. Rev. B}\
  }\textbf {\bibinfo {volume} {49}},\ \bibinfo {pages} {12330} (\bibinfo {year}
  {1994})}\BibitemShut {NoStop}%
\bibitem [{\citenamefont {Ueno}\ \emph {et~al.}(2013)\citenamefont {Ueno},
  \citenamefont {Yamakage}, \citenamefont {Tanaka},\ and\ \citenamefont
  {Sato}}]{25}%
  \BibitemOpen
  \bibfield  {author} {\bibinfo {author} {\bibfnamefont {Y.}~\bibnamefont
  {Ueno}}, \bibinfo {author} {\bibfnamefont {A.}~\bibnamefont {Yamakage}},
  \bibinfo {author} {\bibfnamefont {Y.}~\bibnamefont {Tanaka}}, \ and\ \bibinfo
  {author} {\bibfnamefont {M.}~\bibnamefont {Sato}},\ }\href@noop {} {\bibfield
   {journal} {\bibinfo  {journal} {Phys. Rev. Lett.}\ }\textbf {\bibinfo
  {volume} {111}},\ \bibinfo {pages} {087002} (\bibinfo {year}
  {2013})}\BibitemShut {NoStop}%
\bibitem [{\citenamefont {Asano}\ \emph {et~al.}(2003)\citenamefont {Asano},
  \citenamefont {Tanaka}, \citenamefont {Sigrist},\ and\ \citenamefont
  {Kashiwaya}}]{26}%
  \BibitemOpen
  \bibfield  {author} {\bibinfo {author} {\bibfnamefont {Y.}~\bibnamefont
  {Asano}}, \bibinfo {author} {\bibfnamefont {Y.}~\bibnamefont {Tanaka}},
  \bibinfo {author} {\bibfnamefont {M.}~\bibnamefont {Sigrist}}, \ and\
  \bibinfo {author} {\bibfnamefont {S.}~\bibnamefont {Kashiwaya}},\ }\href@noop
  {} {\bibfield  {journal} {\bibinfo  {journal} {Phys. Rev. B}\ }\textbf
  {\bibinfo {volume} {67}},\ \bibinfo {pages} {184505} (\bibinfo {year}
  {2003})}\BibitemShut {NoStop}%
\bibitem [{\citenamefont {Yada}\ \emph {et~al.}(2014)\citenamefont {Yada},
  \citenamefont {Golubov}, \citenamefont {Tanaka},\ and\ \citenamefont
  {Kashiwaya}}]{49}%
  \BibitemOpen
  \bibfield  {author} {\bibinfo {author} {\bibfnamefont {K.}~\bibnamefont
  {Yada}}, \bibinfo {author} {\bibfnamefont {A.~A.}\ \bibnamefont {Golubov}},
  \bibinfo {author} {\bibfnamefont {Y.}~\bibnamefont {Tanaka}}, \ and\ \bibinfo
  {author} {\bibfnamefont {S.}~\bibnamefont {Kashiwaya}},\ }\href@noop {}
  {\bibfield  {journal} {\bibinfo  {journal} {J. Phys. Soc. Jpn.}\ }\textbf
  {\bibinfo {volume} {83}},\ \bibinfo {pages} {074706} (\bibinfo {year}
  {2014})}\BibitemShut {NoStop}%
\bibitem [{\citenamefont {Kawai}\ \emph {et~al.}(2017)\citenamefont {Kawai},
  \citenamefont {Yada}, \citenamefont {Tanaka}, \citenamefont {Asano},
  \citenamefont {Golubov},\ and\ \citenamefont {Kashiwaya}}]{27}%
  \BibitemOpen
  \bibfield  {author} {\bibinfo {author} {\bibfnamefont {K.}~\bibnamefont
  {Kawai}}, \bibinfo {author} {\bibfnamefont {K.}~\bibnamefont {Yada}},
  \bibinfo {author} {\bibfnamefont {Y.}~\bibnamefont {Tanaka}}, \bibinfo
  {author} {\bibfnamefont {Y.}~\bibnamefont {Asano}}, \bibinfo {author}
  {\bibfnamefont {A.~A.}\ \bibnamefont {Golubov}}, \ and\ \bibinfo {author}
  {\bibfnamefont {S.}~\bibnamefont {Kashiwaya}},\ }\href@noop {} {\bibfield
  {journal} {\bibinfo  {journal} {Phys. Rev. B}\ }\textbf {\bibinfo {volume}
  {95}},\ \bibinfo {pages} {174518} (\bibinfo {year} {2017})}\BibitemShut
  {NoStop}%
\bibitem [{\citenamefont {Saitoh}\ \emph {et~al.}(2012)\citenamefont {Saitoh},
  \citenamefont {Kashiwaya}, \citenamefont {Kashiwaya}, \citenamefont
  {Koyanagi}, \citenamefont {Mawatari}, \citenamefont {Tanaka},\ and\
  \citenamefont {Maeno}}]{28}%
  \BibitemOpen
  \bibfield  {author} {\bibinfo {author} {\bibfnamefont {K.}~\bibnamefont
  {Saitoh}}, \bibinfo {author} {\bibfnamefont {S.}~\bibnamefont {Kashiwaya}},
  \bibinfo {author} {\bibfnamefont {H.}~\bibnamefont {Kashiwaya}}, \bibinfo
  {author} {\bibfnamefont {M.}~\bibnamefont {Koyanagi}}, \bibinfo {author}
  {\bibfnamefont {Y.}~\bibnamefont {Mawatari}}, \bibinfo {author}
  {\bibfnamefont {T.}~\bibnamefont {Tanaka}}, \ and\ \bibinfo {author}
  {\bibfnamefont {Y.}~\bibnamefont {Maeno}},\ }\href@noop {} {\bibfield
  {journal} {\bibinfo  {journal} {Appl. Phys. Express}\ }\textbf {\bibinfo
  {volume} {5}},\ \bibinfo {pages} {113101} (\bibinfo {year}
  {2012})}\BibitemShut {NoStop}%
\bibitem [{\citenamefont {Mao}\ \emph {et~al.}(2000)\citenamefont {Mao},
  \citenamefont {Maeno},\ and\ \citenamefont {Fukazawa}}]{31}%
  \BibitemOpen
  \bibfield  {author} {\bibinfo {author} {\bibfnamefont {Z.~Q.}\ \bibnamefont
  {Mao}}, \bibinfo {author} {\bibfnamefont {Y.}~\bibnamefont {Maeno}}, \ and\
  \bibinfo {author} {\bibfnamefont {H.}~\bibnamefont {Fukazawa}},\ }\href@noop
  {} {\bibfield  {journal} {\bibinfo  {journal} {Mater. Res. Bull.}\ }\textbf
  {\bibinfo {volume} {35}},\ \bibinfo {pages} {1813} (\bibinfo {year}
  {2000})}\BibitemShut {NoStop}%
\bibitem [{32()}]{32}%
  \BibitemOpen
  \href@noop {} {\bibinfo  {journal} {Amumetal 4K (Amuneal Co. Ltd.)}\
  }\BibitemShut {NoStop}%
\bibitem [{\citenamefont {Bouhon}\ and\ \citenamefont {Sigrist}(2010)}]{34}%
  \BibitemOpen
\bibfield  {journal} {  }\bibfield  {author} {\bibinfo {author} {\bibfnamefont
  {A.}~\bibnamefont {Bouhon}}\ and\ \bibinfo {author} {\bibfnamefont
  {M.}~\bibnamefont {Sigrist}},\ }\href@noop {} {\bibfield  {journal} {\bibinfo
   {journal} {New J. Phys.}\ }\textbf {\bibinfo {volume} {12}},\ \bibinfo
  {pages} {043031} (\bibinfo {year} {2010})}\BibitemShut {NoStop}%
\bibitem [{\citenamefont {Asano}\ \emph {et~al.}(2005)\citenamefont {Asano},
  \citenamefont {Tanaka}, \citenamefont {Sigrist},\ and\ \citenamefont
  {Kashiwaya}}]{35}%
  \BibitemOpen
  \bibfield  {author} {\bibinfo {author} {\bibfnamefont {Y.}~\bibnamefont
  {Asano}}, \bibinfo {author} {\bibfnamefont {Y.}~\bibnamefont {Tanaka}},
  \bibinfo {author} {\bibfnamefont {M.}~\bibnamefont {Sigrist}}, \ and\
  \bibinfo {author} {\bibfnamefont {S.}~\bibnamefont {Kashiwaya}},\ }\href@noop
  {} {\bibfield  {journal} {\bibinfo  {journal} {Phys. Rev. B}\ }\textbf
  {\bibinfo {volume} {71}},\ \bibinfo {pages} {214501} (\bibinfo {year}
  {2005})}\BibitemShut {NoStop}%
\bibitem [{\citenamefont {Anwar}\ \emph {et~al.}(2017)\citenamefont {Anwar},
  \citenamefont {Ishiguro}, \citenamefont {Nakamura}, \citenamefont {Yakabe},
  \citenamefont {Yonezawa}, \citenamefont {Takayanagi},\ and\ \citenamefont
  {Maeno}}]{30}%
  \BibitemOpen
  \bibfield  {author} {\bibinfo {author} {\bibfnamefont {M.~S.}\ \bibnamefont
  {Anwar}}, \bibinfo {author} {\bibfnamefont {R.}~\bibnamefont {Ishiguro}},
  \bibinfo {author} {\bibfnamefont {T.}~\bibnamefont {Nakamura}}, \bibinfo
  {author} {\bibfnamefont {M.}~\bibnamefont {Yakabe}}, \bibinfo {author}
  {\bibfnamefont {S.}~\bibnamefont {Yonezawa}}, \bibinfo {author}
  {\bibfnamefont {H.}~\bibnamefont {Takayanagi}}, \ and\ \bibinfo {author}
  {\bibfnamefont {Y.}~\bibnamefont {Maeno}},\ }\href@noop {} {\bibfield
  {journal} {\bibinfo  {journal} {Phys. Rev. B}\ }\textbf {\bibinfo {volume}
  {95}},\ \bibinfo {pages} {224509} (\bibinfo {year} {2017})}\BibitemShut
  {NoStop}%
\bibitem [{\citenamefont {Kambara}\ \emph {et~al.}(2008)\citenamefont
  {Kambara}, \citenamefont {Kashiwaya}, \citenamefont {Yaguchi}, \citenamefont
  {Asano}, \citenamefont {Tanaka},\ and\ \citenamefont {Maeno}}]{36}%
  \BibitemOpen
  \bibfield  {author} {\bibinfo {author} {\bibfnamefont {H.}~\bibnamefont
  {Kambara}}, \bibinfo {author} {\bibfnamefont {S.}~\bibnamefont {Kashiwaya}},
  \bibinfo {author} {\bibfnamefont {H.}~\bibnamefont {Yaguchi}}, \bibinfo
  {author} {\bibfnamefont {Y.}~\bibnamefont {Asano}}, \bibinfo {author}
  {\bibfnamefont {Y.}~\bibnamefont {Tanaka}}, \ and\ \bibinfo {author}
  {\bibfnamefont {Y.}~\bibnamefont {Maeno}},\ }\href@noop {} {\bibfield
  {journal} {\bibinfo  {journal} {Phys. Rev. Lett.}\ }\textbf {\bibinfo
  {volume} {101}},\ \bibinfo {pages} {267003} (\bibinfo {year}
  {2008})}\BibitemShut {NoStop}%
\bibitem [{\citenamefont {Kambara}\ \emph {et~al.}(2010)\citenamefont
  {Kambara}, \citenamefont {Matsumoto}, \citenamefont {Kashiwaya},
  \citenamefont {Kashiwaya}, \citenamefont {Yaguchi}, \citenamefont {Asano},
  \citenamefont {Tanaka},\ and\ \citenamefont {Maeno}}]{37}%
  \BibitemOpen
  \bibfield  {author} {\bibinfo {author} {\bibfnamefont {H.}~\bibnamefont
  {Kambara}}, \bibinfo {author} {\bibfnamefont {T.}~\bibnamefont {Matsumoto}},
  \bibinfo {author} {\bibfnamefont {H.}~\bibnamefont {Kashiwaya}}, \bibinfo
  {author} {\bibfnamefont {S.}~\bibnamefont {Kashiwaya}}, \bibinfo {author}
  {\bibfnamefont {H.}~\bibnamefont {Yaguchi}}, \bibinfo {author} {\bibfnamefont
  {Y.}~\bibnamefont {Asano}}, \bibinfo {author} {\bibfnamefont
  {Y.}~\bibnamefont {Tanaka}}, \ and\ \bibinfo {author} {\bibfnamefont
  {Y.}~\bibnamefont {Maeno}},\ }\href@noop {} {\bibfield  {journal} {\bibinfo
  {journal} {J. Phys. Soc. Jpn.}\ }\textbf {\bibinfo {volume} {79}},\ \bibinfo
  {pages} {074708} (\bibinfo {year} {2010})}\BibitemShut {NoStop}%
\bibitem [{\citenamefont {Ambegaokar}\ and\ \citenamefont
  {Baratoff}(1963)}]{29}%
  \BibitemOpen
  \bibfield  {author} {\bibinfo {author} {\bibfnamefont {V.}~\bibnamefont
  {Ambegaokar}}\ and\ \bibinfo {author} {\bibfnamefont {A.}~\bibnamefont
  {Baratoff}},\ }\href@noop {} {\bibfield  {journal} {\bibinfo  {journal}
  {Phys. Rev. Lett.}\ }\textbf {\bibinfo {volume} {10}},\ \bibinfo {pages}
  {486} (\bibinfo {year} {1963})}\BibitemShut {NoStop}%
\bibitem [{\citenamefont {Barone}\ and\ \citenamefont {Paterno}(1982)}]{33}%
  \BibitemOpen
  \bibfield  {author} {\bibinfo {author} {\bibfnamefont {A.}~\bibnamefont
  {Barone}}\ and\ \bibinfo {author} {\bibfnamefont {G.}~\bibnamefont
  {Paterno}},\ }\enquote {\bibinfo {title} {Physics and applications of the
  josephson effect},}\ \ (\bibinfo  {publisher} {Wiley, New York},\ \bibinfo
  {year} {1982})\BibitemShut {NoStop}%
\bibitem [{\citenamefont {Luke}\ \emph {et~al.}(2000)\citenamefont {Luke},
  \citenamefont {Fudamoto}, \citenamefont {Kojima}, \citenamefont {Larkin},
  \citenamefont {Nachumi}, \citenamefont {Uemura}, \citenamefont {Sonier},
  \citenamefont {Maeno}, \citenamefont {Mao},\ and\ \citenamefont
  {Mori}}]{37-2}%
  \BibitemOpen
  \bibfield  {author} {\bibinfo {author} {\bibfnamefont {G.~M.}\ \bibnamefont
  {Luke}}, \bibinfo {author} {\bibfnamefont {Y.}~\bibnamefont {Fudamoto}},
  \bibinfo {author} {\bibfnamefont {K.~M.}\ \bibnamefont {Kojima}}, \bibinfo
  {author} {\bibfnamefont {M.~I.}\ \bibnamefont {Larkin}}, \bibinfo {author}
  {\bibfnamefont {B.}~\bibnamefont {Nachumi}}, \bibinfo {author} {\bibfnamefont
  {Y.~J.}\ \bibnamefont {Uemura}}, \bibinfo {author} {\bibfnamefont {J.~E.}\
  \bibnamefont {Sonier}}, \bibinfo {author} {\bibfnamefont {Y.}~\bibnamefont
  {Maeno}}, \bibinfo {author} {\bibfnamefont {Z.~Q.}\ \bibnamefont {Mao}}, \
  and\ \bibinfo {author} {\bibfnamefont {Y.}~\bibnamefont {Mori}},\ }\href@noop
  {} {\bibfield  {journal} {\bibinfo  {journal} {Physica B}\ }\textbf {\bibinfo
  {volume} {289-290}},\ \bibinfo {pages} {373} (\bibinfo {year}
  {2000})}\BibitemShut {NoStop}%
\bibitem [{\citenamefont {Goryo}(2008)}]{38}%
  \BibitemOpen
  \bibfield  {author} {\bibinfo {author} {\bibfnamefont {J.}~\bibnamefont
  {Goryo}},\ }\href@noop {} {\bibfield  {journal} {\bibinfo  {journal} {Phys.
  Rev. B}\ }\textbf {\bibinfo {volume} {78}},\ \bibinfo {pages} {060501(R)}
  (\bibinfo {year} {2008})}\BibitemShut {NoStop}%
\bibitem [{\citenamefont {Sakuma}\ \emph {et~al.}(2017)\citenamefont {Sakuma},
  \citenamefont {Nago}, \citenamefont {Ishiguro}, \citenamefont {Kashiwaya},
  \citenamefont {Nomura}, \citenamefont {Kono}, \citenamefont {Maeno},\ and\
  \citenamefont {Takayanagi}}]{39}%
  \BibitemOpen
  \bibfield  {author} {\bibinfo {author} {\bibfnamefont {D.}~\bibnamefont
  {Sakuma}}, \bibinfo {author} {\bibfnamefont {Y.}~\bibnamefont {Nago}},
  \bibinfo {author} {\bibfnamefont {R.}~\bibnamefont {Ishiguro}}, \bibinfo
  {author} {\bibfnamefont {S.}~\bibnamefont {Kashiwaya}}, \bibinfo {author}
  {\bibfnamefont {S.}~\bibnamefont {Nomura}}, \bibinfo {author} {\bibfnamefont
  {K.}~\bibnamefont {Kono}}, \bibinfo {author} {\bibfnamefont {Y.}~\bibnamefont
  {Maeno}}, \ and\ \bibinfo {author} {\bibfnamefont {H.}~\bibnamefont
  {Takayanagi}},\ }\href@noop {} {\bibfield  {journal} {\bibinfo  {journal} {J.
  Phys. Soc. Jpn.}\ }\textbf {\bibinfo {volume} {86}},\ \bibinfo {pages}
  {114708} (\bibinfo {year} {2017})}\BibitemShut {NoStop}%
\bibitem [{\citenamefont {Neils}\ and\ \citenamefont
  {VanHarlingen}(2002)}]{46}%
  \BibitemOpen
  \bibfield  {author} {\bibinfo {author} {\bibfnamefont {W.~K.}\ \bibnamefont
  {Neils}}\ and\ \bibinfo {author} {\bibfnamefont {D.~J.}\ \bibnamefont
  {VanHarlingen}},\ }\href@noop {} {\bibfield  {journal} {\bibinfo  {journal}
  {Phys. Rev. Lett.}\ }\textbf {\bibinfo {volume} {88}},\ \bibinfo {pages}
  {047001} (\bibinfo {year} {2002})}\BibitemShut {NoStop}%
\bibitem [{\citenamefont {Jang}\ \emph {et~al.}(2011)\citenamefont {Jang},
  \citenamefont {Ferguson}, \citenamefont {Vakaryuk}, \citenamefont {Budakian},
  \citenamefont {Chung}, \citenamefont {Goldbart},\ and\ \citenamefont
  {Maeno}}]{51}%
  \BibitemOpen
  \bibfield  {author} {\bibinfo {author} {\bibfnamefont {J.}~\bibnamefont
  {Jang}}, \bibinfo {author} {\bibfnamefont {D.~G.}\ \bibnamefont {Ferguson}},
  \bibinfo {author} {\bibfnamefont {V.}~\bibnamefont {Vakaryuk}}, \bibinfo
  {author} {\bibfnamefont {R.}~\bibnamefont {Budakian}}, \bibinfo {author}
  {\bibfnamefont {S.~B.}\ \bibnamefont {Chung}}, \bibinfo {author}
  {\bibfnamefont {P.~M.}\ \bibnamefont {Goldbart}}, \ and\ \bibinfo {author}
  {\bibfnamefont {Y.}~\bibnamefont {Maeno}},\ }\href@noop {} {\bibfield
  {journal} {\bibinfo  {journal} {Science}\ }\textbf {\bibinfo {volume}
  {331}},\ \bibinfo {pages} {186} (\bibinfo {year} {2011})}\BibitemShut
  {NoStop}%
\bibitem [{\citenamefont {Yasui}\ \emph {et~al.}(2017)\citenamefont {Yasui},
  \citenamefont {Lahabi}, \citenamefont {Anwar}, \citenamefont {Nakamura},
  \citenamefont {Yonezawa}, \citenamefont {Terashima}, \citenamefont {Aarts},\
  and\ \citenamefont {Maeno}}]{52}%
  \BibitemOpen
  \bibfield  {author} {\bibinfo {author} {\bibfnamefont {Y.}~\bibnamefont
  {Yasui}}, \bibinfo {author} {\bibfnamefont {K.}~\bibnamefont {Lahabi}},
  \bibinfo {author} {\bibfnamefont {M.~S.}\ \bibnamefont {Anwar}}, \bibinfo
  {author} {\bibfnamefont {Y.}~\bibnamefont {Nakamura}}, \bibinfo {author}
  {\bibfnamefont {S.}~\bibnamefont {Yonezawa}}, \bibinfo {author}
  {\bibfnamefont {T.}~\bibnamefont {Terashima}}, \bibinfo {author}
  {\bibfnamefont {J.}~\bibnamefont {Aarts}}, \ and\ \bibinfo {author}
  {\bibfnamefont {Y.}~\bibnamefont {Maeno}},\ }\href@noop {} {\bibfield
  {journal} {\bibinfo  {journal} {Phys. Rev. B}\ }\textbf {\bibinfo {volume}
  {96}},\ \bibinfo {pages} {180507(R)} (\bibinfo {year} {2017})}\BibitemShut
  {NoStop}%
\bibitem [{\citenamefont {Hicks}\ \emph {et~al.}(2014)\citenamefont {Hicks},
  \citenamefont {Brodsky}, \citenamefont {Yelland}, \citenamefont {Gibbs},
  \citenamefont {Bruin}, \citenamefont {Barber}, \citenamefont {Edkin},
  \citenamefont {Nishimura}, \citenamefont {Yonezawa}, \citenamefont {Maeno},\
  and\ \citenamefont {Mackenzie}}]{53}%
  \BibitemOpen
  \bibfield  {author} {\bibinfo {author} {\bibfnamefont {C.~W.}\ \bibnamefont
  {Hicks}}, \bibinfo {author} {\bibfnamefont {D.~O.}\ \bibnamefont {Brodsky}},
  \bibinfo {author} {\bibfnamefont {E.~A.}\ \bibnamefont {Yelland}}, \bibinfo
  {author} {\bibfnamefont {A.~S.}\ \bibnamefont {Gibbs}}, \bibinfo {author}
  {\bibfnamefont {J.~A.~N.}\ \bibnamefont {Bruin}}, \bibinfo {author}
  {\bibfnamefont {M.~E.}\ \bibnamefont {Barber}}, \bibinfo {author}
  {\bibfnamefont {S.~D.}\ \bibnamefont {Edkin}}, \bibinfo {author}
  {\bibfnamefont {K.}~\bibnamefont {Nishimura}}, \bibinfo {author}
  {\bibfnamefont {S.}~\bibnamefont {Yonezawa}}, \bibinfo {author}
  {\bibfnamefont {Y.}~\bibnamefont {Maeno}}, \ and\ \bibinfo {author}
  {\bibfnamefont {A.~P.}\ \bibnamefont {Mackenzie}},\ }\href@noop {} {\bibfield
   {journal} {\bibinfo  {journal} {Science}\ }\textbf {\bibinfo {volume}
  {344}},\ \bibinfo {pages} {283} (\bibinfo {year} {2014})}\BibitemShut
  {NoStop}%
\bibitem [{\citenamefont {Hassinger}\ \emph {et~al.}(2017)\citenamefont
  {Hassinger}, \citenamefont {Bourgeois-Hope}, \citenamefont {Taniguchi},
  \citenamefont {RenedeCotret}, \citenamefont {Grissonnanche}, \citenamefont
  {Anwar}, \citenamefont {Maeno}, \citenamefont {Doiron-Leyraud},\ and\
  \citenamefont {Taillefer}}]{54}%
  \BibitemOpen
  \bibfield  {author} {\bibinfo {author} {\bibfnamefont {E.}~\bibnamefont
  {Hassinger}}, \bibinfo {author} {\bibfnamefont {P.}~\bibnamefont
  {Bourgeois-Hope}}, \bibinfo {author} {\bibfnamefont {H.}~\bibnamefont
  {Taniguchi}}, \bibinfo {author} {\bibfnamefont {S.}~\bibnamefont
  {RenedeCotret}}, \bibinfo {author} {\bibfnamefont {G.}~\bibnamefont
  {Grissonnanche}}, \bibinfo {author} {\bibfnamefont {M.~S.}\ \bibnamefont
  {Anwar}}, \bibinfo {author} {\bibfnamefont {Y.}~\bibnamefont {Maeno}},
  \bibinfo {author} {\bibfnamefont {N.}~\bibnamefont {Doiron-Leyraud}}, \ and\
  \bibinfo {author} {\bibfnamefont {L.}~\bibnamefont {Taillefer}},\ }\href@noop
  {} {\bibfield  {journal} {\bibinfo  {journal} {Phys. Rev. X}\ }\textbf
  {\bibinfo {volume} {7}},\ \bibinfo {pages} {011032} (\bibinfo {year}
  {2017})}\BibitemShut {NoStop}%
\bibitem [{\citenamefont {Pustogow}\ \emph {et~al.}()\citenamefont {Pustogow},
  \citenamefont {Juo}, \citenamefont {Chronister}, \citenamefont {Su},
  \citenamefont {Sokolov}, \citenamefont {Jerzembeck}, \citenamefont
  {Mackenzie}, \citenamefont {Hicks}, \citenamefont {Nikugawa}, \citenamefont
  {Raghu}, \citenamefont {Bauer},\ and\ \citenamefont {Brown}}]{63}%
  \BibitemOpen
  \bibfield  {author} {\bibinfo {author} {\bibfnamefont {A.}~\bibnamefont
  {Pustogow}}, \bibinfo {author} {\bibfnamefont {Y.}~\bibnamefont {Juo}},
  \bibinfo {author} {\bibfnamefont {A.}~\bibnamefont {Chronister}}, \bibinfo
  {author} {\bibfnamefont {Y.~S.}\ \bibnamefont {Su}}, \bibinfo {author}
  {\bibfnamefont {S.}~\bibnamefont {Sokolov}}, \bibinfo {author} {\bibfnamefont
  {F.}~\bibnamefont {Jerzembeck}}, \bibinfo {author} {\bibfnamefont {A.~P.}\
  \bibnamefont {Mackenzie}}, \bibinfo {author} {\bibfnamefont {C.~W.}\
  \bibnamefont {Hicks}}, \bibinfo {author} {\bibfnamefont {N.}~\bibnamefont
  {Nikugawa}}, \bibinfo {author} {\bibfnamefont {S.}~\bibnamefont {Raghu}},
  \bibinfo {author} {\bibfnamefont {E.~D.}\ \bibnamefont {Bauer}}, \ and\
  \bibinfo {author} {\bibfnamefont {S.~E.}\ \bibnamefont {Brown}},\ }\href@noop
  {} {\bibfield  {journal} {\bibinfo  {journal} {arXiv:1904:00047}\ }}\bibinfo
  {note} {During the preparation of the present paper, we noticed this
  paper.}\BibitemShut {Stop}%
\bibitem [{\citenamefont {Qi}\ \emph {et~al.}(2009)\citenamefont {Qi},
  \citenamefont {Hughes}, \citenamefont {Raghu},\ and\ \citenamefont
  {Zhang}}]{57}%
  \BibitemOpen
  \bibfield  {author} {\bibinfo {author} {\bibfnamefont {X.-L.}\ \bibnamefont
  {Qi}}, \bibinfo {author} {\bibfnamefont {T.~L.}\ \bibnamefont {Hughes}},
  \bibinfo {author} {\bibfnamefont {S.}~\bibnamefont {Raghu}}, \ and\ \bibinfo
  {author} {\bibfnamefont {S.-C.}\ \bibnamefont {Zhang}},\ }\href@noop {}
  {\bibfield  {journal} {\bibinfo  {journal} {Phys. Rev. Lett.}\ }\textbf
  {\bibinfo {volume} {102}},\ \bibinfo {pages} {187001} (\bibinfo {year}
  {2009})}\BibitemShut {NoStop}%
\bibitem [{\citenamefont {Sato}(2009)}]{58}%
  \BibitemOpen
  \bibfield  {author} {\bibinfo {author} {\bibfnamefont {M.}~\bibnamefont
  {Sato}},\ }\href@noop {} {\bibfield  {journal} {\bibinfo  {journal} {Phys.
  Rev. B}\ }\textbf {\bibinfo {volume} {79}},\ \bibinfo {pages} {214526}
  (\bibinfo {year} {2009})}\BibitemShut {NoStop}%
\bibitem [{\citenamefont {Sato}(2010)}]{58-2}%
  \BibitemOpen
  \bibfield  {author} {\bibinfo {author} {\bibfnamefont {M.}~\bibnamefont
  {Sato}},\ }\href@noop {} {\bibfield  {journal} {\bibinfo  {journal} {Phys.
  Rev. B}\ }\textbf {\bibinfo {volume} {81}},\ \bibinfo {pages} {220504(R)}
  (\bibinfo {year} {2010})}\BibitemShut {NoStop}%
\bibitem [{\citenamefont {Fu}(2011)}]{59}%
  \BibitemOpen
  \bibfield  {author} {\bibinfo {author} {\bibfnamefont {L.}~\bibnamefont
  {Fu}},\ }\href@noop {} {\bibfield  {journal} {\bibinfo  {journal} {Phys. Rev.
  Lett.}\ }\textbf {\bibinfo {volume} {106}},\ \bibinfo {pages} {106802}
  (\bibinfo {year} {2011})}\BibitemShut {NoStop}%
\bibitem [{\citenamefont {M.~Sato}(2014)}]{60}%
  \BibitemOpen
  \bibfield  {author} {\bibinfo {author} {\bibfnamefont {T.~M.}\ \bibnamefont
  {M.~Sato}, \bibfnamefont {A.~Yamakage}},\ }\href@noop {} {\bibfield
  {journal} {\bibinfo  {journal} {Physica E}\ }\textbf {\bibinfo {volume}
  {55}},\ \bibinfo {pages} {20} (\bibinfo {year} {2014})}\BibitemShut {NoStop}%
\bibitem [{\citenamefont {Ivanov}(2001)}]{61}%
  \BibitemOpen
  \bibfield  {author} {\bibinfo {author} {\bibfnamefont {D.~A.}\ \bibnamefont
  {Ivanov}},\ }\href@noop {} {\bibfield  {journal} {\bibinfo  {journal} {Phys.
  Rev. Lett.}\ }\textbf {\bibinfo {volume} {86}},\ \bibinfo {pages} {268}
  (\bibinfo {year} {2001})}\BibitemShut {NoStop}%
\bibitem [{\citenamefont {Nayak}\ \emph {et~al.}(2008)\citenamefont {Nayak},
  \citenamefont {Simon}, \citenamefont {Stern}, \citenamefont {Freedman},\ and\
  \citenamefont {DasSarma}}]{62}%
  \BibitemOpen
  \bibfield  {author} {\bibinfo {author} {\bibfnamefont {C.}~\bibnamefont
  {Nayak}}, \bibinfo {author} {\bibfnamefont {S.~H.}\ \bibnamefont {Simon}},
  \bibinfo {author} {\bibfnamefont {A.}~\bibnamefont {Stern}}, \bibinfo
  {author} {\bibfnamefont {M.}~\bibnamefont {Freedman}}, \ and\ \bibinfo
  {author} {\bibfnamefont {S.}~\bibnamefont {DasSarma}},\ }\href@noop {}
  {\bibfield  {journal} {\bibinfo  {journal} {Rev. Mod. Phys.}\ }\textbf
  {\bibinfo {volume} {80}},\ \bibinfo {pages} {1083} (\bibinfo {year}
  {2008})}\BibitemShut {NoStop}%
\end{thebibliography}%

\end{document}